\documentclass{pasj01}
\usepackage{color}

\newcommand{\sgr}{SGR~1806$-$20}    
\newcommand{\oneE}{1E~1547.0$-$5408}    
\newcommand{\NuS}{NuSTAR}
\newcommand{\ASCA}{ASCA}
\newcommand{\Su}{Suzaku}
\newcommand{\Bd}{B_{\rm d}}
\newcommand{\Bt}{B_{\rm t}}
\newcommand{\Ppr}{P_{\rm pr}}
\newcommand{\Prot}{P_{\rm rot}}
\newcommand{\Pch}{{\cal P}_{\rm ch}}
\newcommand{\zz}{Z_{4}^{2}}
\newcommand{\Tpk}{T_{16.4}}

\begin{document}

\SetRunningHead{Makishima, K.}{Free precession in SGR 1806$-$20}

\Received{2023/12/04}
\Accepted{2024/04/18}

\title{Discovery of the Free Precession in the Magnetar  \sgr\  with the ASCA GIS
}


\author{
Kazuo  \textsc{Makishima}\altaffilmark{1,2},
Nagomi \textsc{Uchida}\altaffilmark{3},
and
Teruaki \textsc{Enoto}\altaffilmark{4,5}
}
\email{maxima@phys.s.u-tokyo.ac.jp}

\altaffiltext{1}{
Department of Physics, The University of Tokyo,
7-3-1 Hongo, Bunkyo-ku, Tokyo 113-0033
}
\altaffiltext{2}{
Kavli Institute for the Physics and Mathematics of the Universe (WPI),
The University of Tokyo,
5-1-5 Kashiwa-no-ha, Kashiwa, Chiba, 
277-8683
}
\altaffiltext{3}{
Institute of Space and Astronautical Science, JAXA,
3-1-1 Yoshinodai, Chuo-ku, Sagamihara 252-5210, Japan
}
\altaffiltext{4}{
Department of Physics, Kyoto University,
Kitashirakawa Oiwake-cho, Sakyo-ku, Kyoto, Japan 606-8502
}
\altaffiltext{5}{
Extreme Natural Phenomena RIKEN Hakubi Research Team,
Cluster for Pioneering Research, RIKEN, 2-1 Hirosawa, 
Wako, Saitama, Japan 351-0198
}

\maketitle

\begin{abstract}
Four X-ray data sets  of the Soft Gamma Repeater \sgr,
taken with the Gas Imaging Spectrometer (GIS) onboad \ASCA,
were analyzed.
Three of them were acquired over  1993 October 9--20,
whereas the last one  in 1995 October.
Epoch-folding analysis of the 2.8--12 keV signals 
confirmed the $\sim 7.6$ s pulses in these data,
which \citet{Kouveliotou98} reported as one of 
the earliest pulse detections from this object.
In the 1995 observation, 3--12 keV pulses were 
phase modulated with a period of $T =16.4 \pm 0.4$ ks,
and an amplitude of $\sim 1$ s.
This makes a fourth example of the behavior observed from magnetars.
Like in  the previous three sources,
the pulse-phase modulation of \sgr\ disappeared at $\lesssim 2.5$ keV,
where the soft X-ray component dominates.
In the 1993 data sets, this periodic modulation  was reconfirmed,
and successfully phase-connected coherently across  the 11 d interval.
As a result, the modulation period was refined to  $T =16.435 \pm 0.024$ ks.
The implied high stability of the phenomenon strengthens its
interpretation in terms of free precession of the neutron star,
which is deformed to an asphericity of $\sim 10^{-4}$,
presumably by the stress of toroidal magnetic fields reaching $\sim 10^{16}$ G.
Toroidal fields of this level can be 
common among magnetars.
\end{abstract}
%

\section{INTRODUCTION}
\label{sec:intro}
Consider an axisymmetric  rigid body with $I_1=I_2 \ne I_3$,
where $I_j$ is the moment of inertia around
the principal axes  $\hat{x}_j~(j=1,2,3)$,
with $\hat{x}_3$ the body's symmetry axis.
When the body is free from external torque,
its angular momentum $\vec L$ is conserved,
and  its dynamics around the center of gravity 
is split into two modes \citep{Landau&Lifshitz}
that degenerate when the body is spherical.
One is {\em free precession},
wherein $\hat{x}_3$ rotates
(as seen from the inertial frame) around $\vec L$
with a constant precession period $\Ppr = 2 \pi I_1/L$,
and a constant  {\em wobbling angle} $\alpha$ 
relative to $\vec L$.
(This should not be confused with {\em forced precession}
that is often observed in a spinning top.)
The other is rotation around $\hat{x_3}$ 
with a rotation period $\Prot=2 \pi I_3/L = \Ppr/(1+\epsilon)$,
where $\epsilon \equiv (I_1-I_3)/I_3$ is {\em asphericity}.
When $\alpha \ne 0$ and  the body's emission is symmetric around $\hat{x}_3$,
we can detect $\Ppr$  as the pulsation,
whereas $\Prot$ is undetectable (see a discussion in subsection \ref{subsubsec:free_precession}).
If $\alpha \ne 0$ and the emission violates the symmetry around $\hat{x}_3$,
the phase of the pulsation at $\Ppr$ becomes modulated
at the  beat period between $\Ppr$ and $P_{\rm rot}$,
given as
\begin{equation}
T = \frac{\Ppr}{\epsilon \cos \alpha} 
= \frac{1}{\cos \alpha} \left(P_{\rm rot}^{-1} -{\Ppr}^{-1} \right)^{-1}
\label{eq:slip_period}
\end{equation}
 \citep{Butikov06}. 
 This {\it pulse-phase modulation} (PPM) provides
evidence for the free precession
in an  asymmetrically  radiating celestial object
that is axially deformed ($\epsilon \ne 0$) and has $\alpha \ne 0$.

Although astrophysical examples of free precession remained relatively limited,
we have detected its evidence from three  magnetars;
4U 0142+61 (\cite{Makishima14}; 2019),
\oneE\ (\cite{Makishima16}; 2021a),
and SGR~1900+14 \citep{Makishima21b}.
In these objects, the  hard X-ray pulses with a period $P= P_{\rm pr}$ 
were found to exhibit the PPM effect,
with a long period of  $T\sim 10^4 P$
which can be identified with $T$ in equation~(\ref{eq:slip_period}).
Further  assuming $\cos \alpha \sim 1$, 
we find that these neutron stars (NSs) are 
deformed to $\epsilon \sim P/T \sim 10^{-4}$,
and performs free precession.

Since the centrifugal effect is  much smaller 
(estimated to be $\epsilon \sim 10^{-7}$) in these slowly rotating NSs,
the deformation must be due to magnetic stress.
Then, the inferred magnetic field becomes $B\sim 10^{16}$ G,
when combined with a theoretical prediction \citep{Ioka&Sasaki04}  as
\begin{equation}
\epsilon \sim 10^{-4}(B/10^{16}~{\rm G})^2.
\label{eq:Bt_epsilon}
\end{equation}
Because this $B$ is much higher than the dipole magnetic fields
of these objects, $\Bd = (1-7) \times 10^{14}$ G,
the magnetic fields that cause the deformation are
considered to be confined inside these NSs,
in the form of toroidal magnetic fields, $\Bt$.

To reinforce this scenario,  we study \sgr,
with the primary aim to search for the  PPM effect.
If this phenomenon is common to magnetars,
it should also be detected from \sgr,
the prototypical object
which  has connected the two apparently unrelated astrophysical concepts,
Soft Gamma Repeaters and magnetars (e.g., \cite{Mereghetti08}).

In all the preceding three objects (e.g., \cite{Makishima16}),
the PPM  was observed  in their spectral hard X-ray component (HXC),
but  absent in their soft X-ray component (SXC).
A likely interpretation is that the SXC is emitted symmetrically
around  $\hat{x}_3$ (which we identify with the magnetic axis),
whereas the HXC breaks that symmetry.
Among the known magnetars, \sgr\ hosts by far the brightest HXC
that  extends down to $\sim 3$ keV
\citep{Enoto10}.
Therefore, the PPM in \sgr, if any, 
should be detected down  to $\sim 3$ keV,
and would disappear at lower energies.
To  confirm this conjecture  makes our  second objective.

Since the first X-ray ``identification'' in 1993 (subsection~\ref{subsec:obs_history}),
\sgr\ was observed repeatedly by various X-ray missions.
Considering the  second objective,
we may utilize not only  hard X-ray data, but also those below 10 keV.
We hence select archival 0.7--12 keV data from  \ASCA\ 
acquired in 1993 and 1995, for the following reasons.
(1) These \ASCA\ data provide one of the earliest information
on the {\it persistent emission} from \sgr,
and  can be regarded as a start point of the investigation
of this subject in \sgr.
(2) As detailed later (subsection~\ref{subsec:obs_GIS}),
the Gas Imaging Spectrometer (GIS) onboard \ASCA\
 is ideal to our study.
(3) Because the PPM period $T$ is expected to appear at several tens of kiloseconds,
we need a total data span of  $\gtrsim 50$ ks.
This makes \ASCA\ better suited than, e.g., RXTE or XMM-Newton.
(4) As well known, \sgr\  produced a giant flare (hereafter GF) on 2004 December 27 
(e.g., \cite{GiantFlare}), and since then, it showed 
enhanced timing noise (e.g., \cite{Younes15}), for 10 yrs or more.
This could hamper our timing studies,
but the \ASCA\ data, acquired way before the GF,
is expected to be cleaner in its timing behavior.

So far, we have attributed the PPM in the HXC
to the free precession of a deformed NS.
If this view is correct,
the phase modulation, due to celestial mechanics,
must have high stability.
In our previous studies,
consistent values of $T$ were in fact measured in separate observations
(each  for 1--2 days typically) of the same source,
but they were so widely apart 
that we were unable to coherently phase connect the PPM  across them.
Now, the \ASCA\ data in 1993 were acquired in 3 separate pointings
which span a total  interval of 11 days.
Therefore, if the PPM is detected, 
we can for the first time attempt to coherently phase connect
the periodic  modulation,
across a time span which is an order of magnitude 
longer than was available before.
This is the third objective of the present study.

\section{OBSERVATIONS}
\label{sec:obs}

\subsection{The Gas Imaging Spectrometer (GIS)}
\label{subsec:obs_GIS}

The Gas Imaging Spectrometer (GIS; \cite{GIS1,GIS2}) onboard \ASCA\
consists of a pair of identical imaging gas scintillation proportional counters,
named GIS2 and GIS3.
They are placed at the focal planes of the X-ray Telescope (XRT; \cite{XRT}),
and  cover a wide field of view ($0^\circ.75$ diameter)
with a moderate angular resolution ($\approx 4'$).
The GIS also  realizes a high sensitivity and a low background,
over a broad energy band of  0.7--12 keV
which covers both the HXC and SXC of \sgr.
The modest energy resolution (8\% FHHM at 7 keV)  of the GIS
is sufficient for the present  purpose.
Being gas detectors, the GIS also has a high time resolution 
($\Delta t= 61$ or 488  $\mu$s)
which is sufficient in the present work,
together with a low dead time
which allows detections of burst-like phenomena.

\begin{table*}
\caption{X-ray data sets from the \ASCA\ GIS utilized in the present work.}
\label{tbl:obs}
\begin{center}
\begin{tabular}{llcccccccccc}
\hline
ID & Start date  &MJD $^*$ &  Data span$^\dagger$&  Net exposure &Total photons$^\ddagger$ & Flux$^\S$ \\
\hline \hline
G93pA& 1993  October 9 & 49269.69  &  27.0 ks  & 15.4 ks & 1990&     $1.41^{+0.16}_{-0.15}$   \\
G93pB& 1993 October 10 & 49270.88&  59.3 ks  &  31.7 ks &7529  &   $1.29 ^{+0.09}_{-0.08}$  \\
G93p& (G93pA+G93pB)  &   ---         & 162.1 ks  &  47.1 ks &9519  &  ---  \\
\hline
G93f& 1993 October 20 & 49280.22&   90.9 ks  & 39.7 ks & 9613  &  $1.26  ^{+0.08}_{-0.07}$  \\
\hline
G93T& (G93p+G93f) &  --- &  1000.8 ks  & 86.8 ks & 19132  &   ---  \\
\hline
G95 & 1995 October 16 &50006.07 &  112.4 ks  &58.9 ks & 13556 &  $0.97^{+0.06}_{-0.05}$  \\
\hline
\end{tabular}
\begin{footnotesize}
\begin{itemize}
\item[$^*$] The Modified Julian Date of the first photon in the data.
\item[$^\dagger$] Gross elapsed time of the observation, from the start to the end.
\item[$^\ddagger$]  The total number of 0.7--12 keV events detected with GIS2+GIS3, including background.
\item[$^\S$]  Absorption-removed 2--10 keV fluxes in $10^{-11}$ erg s$^{-1}$ cm$^{-2}$, 
summed over the SXC and the HXC. Errors are 90\% confidence limits.
\end{itemize}
\end{footnotesize}
\end{center}
\end{table*}

\vspace*{-3mm}
\subsection{A brief history}
\label{subsec:obs_history}
We  briefly review the dramatic progress
of the knowledge on \sgr\ that took place in the mid 1990s.
As described in \citet{Murakami94},
the fourth Japanese X-ray satellite \ASCA\  \citep{Tanaka94}
had been in orbit for 7 months
when a  burst activity of \sgr\ was detected,
after nearly a decade,
by the CGRO/BATSE on 1993 September 29.
At that time, the object was only coarsely ($\sim 1^\circ \times 4'$) localized 
by the interplanetary burst-timing triangulation.

On 1993 October 9 and 10,
four short pilot pointings with  \ASCA\ were conducted,
to cover the elongated error region with  the wide field of view of the GIS.
During one of them, onto ``position-A" in \citet{Murakami94},
a short burst took place,
as detected also by the CGRO/BATSE \citep{Kouveliotou94}.
Fortunately, this burst occurred at a periphery of the GIS field of view;
$\sim 170$ signal photons were detected by GIS2+GIS3,
of which about 10 were telemetered to ground
with their full  information (energy, position, and time).
Using  these 10  photons,
the source was  localized  to 
an accuracy of several arcmin.
An associated persistent source was also detected.

In the second of the four pilot pointings,
which was onto ``position-B" of \citep{Murakami94},
the source  happened to be closer to the \ASCA's optical axis.
Although no bursts took place,
the persistent emission was detected again by the GIS,
and also by the co-aligned narrower-field CCD instrument, 
the Solid-State Imaging Spectrometer, or SIS \citep{SIS}.
The source location was refined by the finer position resolution of the SIS,
and was  named AX1805.7$-$2025.
This has provided  the ``X-ray identification" of \sgr.

About 10 days later, on 1993 October 20, 
a follow-up  \ASCA\ pointing was made to reconfirm AX1805.7$-$2025.
The source was observed again with  \ASCA\ in 1995 October.
These follow-up observations lasted longer than the pilot pointings.

\citet{Kouveliotou98}, hereafter KEA98,
combined five RXTE observations of \sgr\ 
made in 1996 November.
They discovered the  source  pulsation at  a period of $P=7.47655$ s,
and verified that the object  is a magnetized NS.
They also returned to the 1993 and 1995 \ASCA\ data,
and confirmed the pulsation at $7.46851 \pm 0.00025$ s 
and $7.4738\pm 0.001$ s, respectively.
Based on the implied rapid spin down rate
$\dot P=0.8 \times 10^{-10}$ s s$^{-1}$ from 1993 to 1996, 
and assuming  the energy loss via magnetic dipole radiation,
they concluded that the NS has a strong dipole field
of $B_{\rm d} = 8 \times 10^{14}$ G.
This made \sgr\  the first magnetar in its genuine sense.
Since then, \sgr\ was observed frequently,
e.g., with RXTE and XMM-Newton
(e.g., \cite{Younes15}),
across the 2004 December GF.

\subsection{\ASCA\ GIS data}
\label{subsec:obs_ASCA_data}

We utilize the  \ASCA\ GIS data,
acquired in 1993 and 1995, 
as described above and detailed  in table~\ref{tbl:obs}.
There, G93pA and G93pB denote two of the four pilot-survey pointings,
at  positions A and  B of \citet{Murakami94},  respectively.
The short burst was detected by the GIS in G93pA,
whereas the source was localized by the SIS in G93pB.
We also  utilize the two follow-up observations,
made in 1993 and 1995, as descried above;
they  are denoted by  G93f  and G95, respectively.
We do not use the SIS data from any of these observations,
because of the  insufficient time resolution (either 2 s or 8 s).

The remaining two pointings  in 1993, 
at ``position-C" and  ``position-D"  \citep{Murakami94},
are not utilized, 
because the source was outside the GIS  field of view.
\sgr\ was within the GIS field of view  also on 1996 April 2, 
but its location was far from the optics axis,
and the data had a short time span of $\approx 20$ ks.
Therefore, this data set is not used, either.

For each observation to be used,
we retrieved screened 0.7--12 keV GIS events
from the JAXA/DARTS Website.
On-source events were extracted from a circular region 
of radius $0^\circ .1$ around the source,
and  those from GIS2 and GIS3 were coadded.
The obtained backgrond-inclusive  photon number is givenin table~\ref{tbl:obs}.
The count rate is twice lower in G93pA than in the others,
because the source was then farther from the XRT optical axis,
where the vignetting effect is severer.

Because G93pA is short with fewer signal photons,
and is separated only by $\sim 1$ d from G93pB,
we merge together G93pA and G93pB
into a single data set denoted as G93p,
and mainly analyze it.
This G93p has a  similar number of photons to G93f,
whereas its total time span, $S=162.1$ ks, 
is nearly twice as long as that  of G93f (S=90.9 ks).
After G93p and G93f are analyzed individually,
they are further merged together into a longer data set  named G93T, 
which is utilized in more detailed timing studies.
The total time span of G93T reaches $S=1000.8$ ks,
although its net exposure is only  8.7\% of this;
the remaining 91.3\% is data gaps, of which the longest one, 
from the end of G93pB to the start of G93f,
is 760.6 ks (8.80 d) long.



\section{DATA ANALYSIS AND RESULTS }
\label{sec:ana}

The GIS data are analyzed without subtracting the background,
which amounts to $\sim 50\%$ of the total on-source counts.
Only when estimating approximate source fluxes listed in table~\ref{tbl:obs},
we subtract a standard background.
The arrival times of these X-ray events are converted 
to those to be measured at the Solar-system barycenter.
The data are not corrected for aspect efficiencies
(collimator transmission and vignetting).
Since the object was burst active in 1993,
we produced a 10-s binned light curve
from each data set using a broad energy range,
and searched those bins where the counts
exceed 2.5 sigmas above the average.
In G93pA, we reconfirmed and removed the 
10 burst photons (bunched in a single bin)
describe in  \citet{Murakami94},
whereas no bursts were detected in the other data sets.

\subsection{Periodograms}
\label{subsec:ana_PG}

In the timing studies below,
we search the photon time-series data sets for periodicities,
using the standard epoch-folding analysis,
combined with the $Z_m^2$ statistics
\citep{LS5039GIS}.
The harmonic number for $Z_m^2$ is chosen as $m=2$ or 4,
 depending on the condition.
The analysis incorporates the spin-down rate measured around 1993-1996,
$\dot P = 0.8 \times 10^{-10}$ s s$^{-1}$ (KEA98).
Its effect is negligible for individual data sets (G93a, G93b, and G95),
because $P$ would change by only $\lesssim 6~\mu$s 
across each observation.
However, when dealing with G93T, 
its effect is  significant.

\begin{figure}
\includegraphics[width=8.2cm]{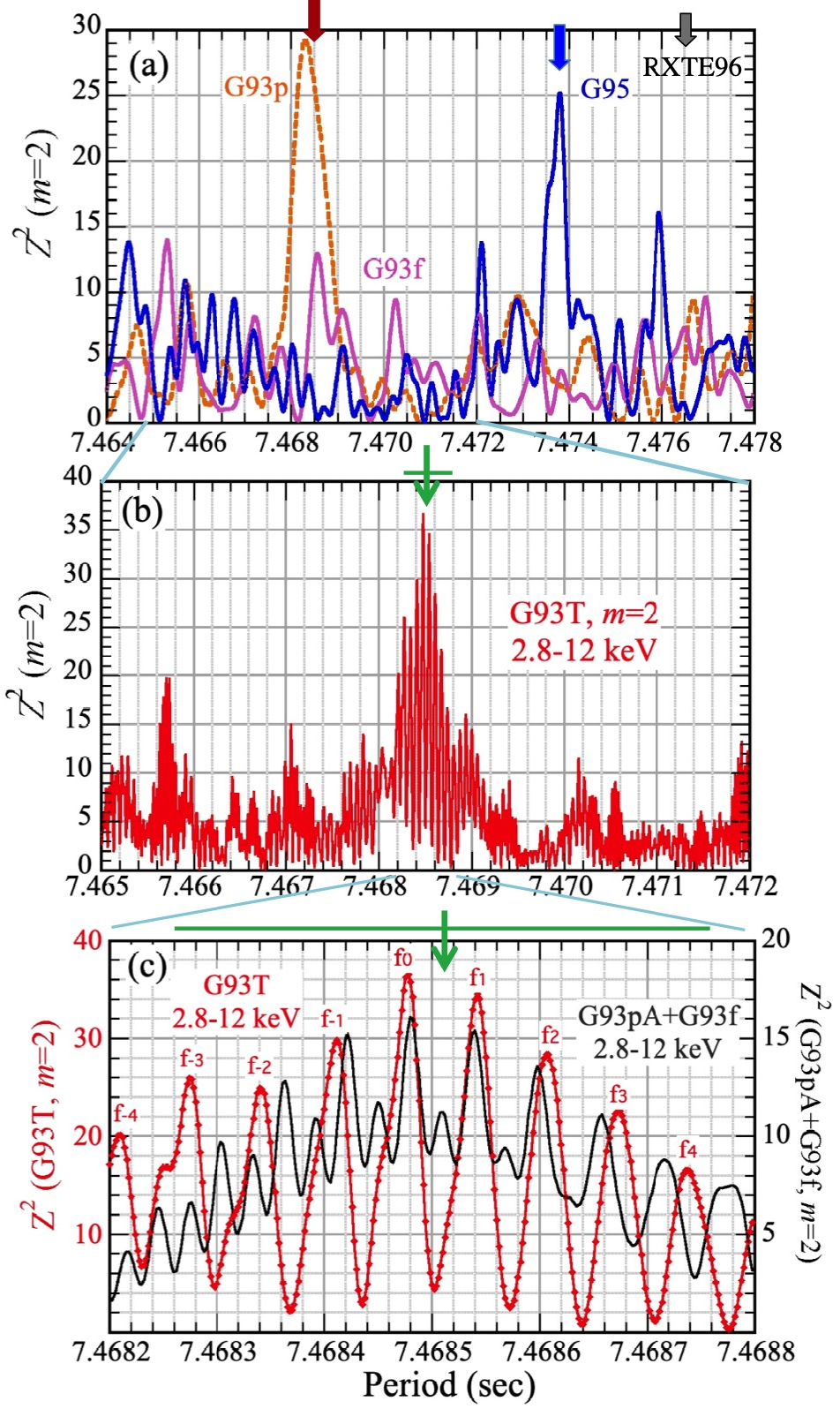}
\caption{Periodograms (PGs) in 2.8--12 keV with $m=2$,
showing the values of  $Z_2^2$  as a function of the assumed period $P$.
(a) PGs from G93p (dashed orange), 
G93f (magenta), and G59 (blue) are shown superposed together.
Measurements by KEA98 are indicated by downward arrows.
(b) A  PG from G93T. 
The result by KEA98 is indicated by a green arrow.
(c) Details of panel (b) around the peak, 
where the fringe numbers are given in red. 
The result from G93pA+G93f is superposed in black (right ordinate).
}
\label{fig:G9395_PGs}
\end{figure}

\begin{table*}
\caption{Results of the periodogram and demodulation analyses of the \ASCA\ GIS data.$^*$}
\begin{center}
\label{tbl:results}
\begin{tabular}{lccclcccccc}
\hline 
Data  &Energy $^\dagger$ & $m$& Condition$^\ddagger$& $P_0$ (s)$^\S$&  $Z_m^2$ & $T$ (ks)  & $A^\|$ & $\psi^{\#}$  \\
\hline \hline 
G93p & 2.8--12& 2& Raw & 7.46831(10)  & 29.2& ---  & --- &  --- \\ 
         & 3--12& 4& Demod. &  7.46836(7)   & 46.1 & $16.7^{+1.5}_{-1.1}$ & 0.6 &  $270^\circ$ \\
 \hline 
 G93f  &2.8--12& 2& Raw    &  7.46856(11)   & 12.9& ---  & --- &  ---\\
         & 3--12& 4& Demod. &   7.46856(7)   & 33.9& $16.5^{+0.5}_{-0.4}$ & 1.0 &  $255^\circ$ \\
 \hline 
G93T  &2.8--12& 2 & Raw &   7.46851(10)$^{**}$ &  36.6  & --- &  --- \\
        & 2.55--12& 4& Demod. &  7.468484(3)  & 63.9& $16.435\pm 0.024$  &1.0 &  $25^\circ$ \\
 \hline 
G95   & 2.8--12& 2& Raw & 7.47380(6)  & 25.2 &  --- & --- &  --- \\
          & 3--12& 4& Demod. &  7.47385(4)  & 54.4 & $16.4\pm 0.4$ & 0.9 &  $235^\circ$ \\
\hline 
\hline
\end{tabular}
\begin{footnotesize}
\begin{itemize}
\setlength{\itemsep}{0mm}
\item[$^{*}$] All assuming a period change rate of $0.8 \times10^{-10}$ s s$^{-1}$.
Errors refer to $68\%$ confidence limits.
\item[$^\dagger$] The utilized energy range in keV.
\item[$^\ddagger$] ``Raw" and ``Demod" mean without and with the demodulation, respectively.
\item[$^\S$] The barycentric pulse period at the start of each observation,
specified by the DeMD peak.
\item[$^\|$] The modulation amplitude $A$  is in units of  s, with a typical error by $\pm 0.2$ s.
\item[$^\#$] $\psi$ has a typical error by $\pm 20^\circ$.
It should take any value in the $0-360^\circ$ range.
\item[$^{**}$] The error range covers four fringes from  f$_{-1}$ to $f_{2}$.
\end{itemize}
\end{footnotesize}
\end{center}
\end{table*}

From  G93p, G93f, and G95,
we produced respective periodograms (hereater PGs)
around the reported pulse periods.
A harmonic number $m=2$  was chosen for $Z_m^2$,
because raw pulse profiles of magnetars are generally double-peaked.
The energy range was set to 2.8.0--12 keV
to approximately remove  the SXC contribution,
because differences between the soft and hard pulse profiles 
sometimes degrade the pulse significance.
The results are shown in figure~\ref{fig:G9395_PGs}a superposed together.
The PGs form G93p and G93f both exhibit a peak
at $P\approx 7.4684$ s, and that of G95 at $P\approx 7.4738$ s.
These periods are summarize in table~\ref{tbl:results},
together with their 68\%-confidence errors
as determined in Appendix A (in particular, subsection A2 therein).
Since these results agree with those which KEA98 derived from the same \ASCA\ data,
we reconfirm the pulse detections and $P$ measurements by KEA98.

The same analysis applied to the merged data G93T,
employing a finer period step of $2~\mu$s,
yielded a PG shown in figure~\ref{fig:G9395_PGs}b.
Its behavior near the peak is expanded in figure~\ref{fig:G9395_PGs}c.
We observe a very clear fine structure,
comprising a series of peaks separated by $\delta P \approx 66~\mu$s.
Since this $\delta P$ and the total data span $S = 1000$ ks of G93T
approximately satisfy the relation $S\times \delta P \approx P^2$,
or $S/P - S/(P+\delta P) \approx 1$,
the fine structure is undoubtedly ``fringes"
produced by interference between the periodicities in G93p and G93f,
across the $\approx 8.8$ d data gap between them.
For convenience, we number the fringes
from $-4$ to +4 as in figure~\ref{fig:G9395_PGs}c,
and denote the $n$-th fringe by f$_n$.
Table ~\ref{tbl:fringe_param} summarizes the parameters of these fringes.

In  figure~\ref{fig:G9395_PGs}c,
the highest fringe  f$_0$, at $P=7.46848$ s,  has $Z_2^2=36.58$.
Since $Z_m^2$ obeys a chi-square distribution of $2m$ degrees of freedom,
the  probability for a $Z_2^2$ value exceeding 36.58 to appear by chance
is  $2.2 \times 10^{-7}$ in a single trial.
On the other hand, against $S=10^{6}$ s,
the minimum and maxim Fourier wave numbers
covered by figure~\ref{fig:G9395_PGs}b are
$10^{6}/7.472=133,832.9$
and $10^{6}/7.465=133,958.5$, respectively,
Their difference, 125.5, represents the number of independent Fourier waves, 
and hence  approximates the effective number of period trials.
Then, the chance probability of the peak in figure~\ref{fig:G9395_PGs}b
is estimated as $\Pch=2.2 \times 10^{-7} \times 125.5=2.8 \times 10^{-5}$.
This is tighter than  the value of $3.6\times 10^{-4}$,
which KEA98 derived using presumably  (not clearly stated in their paper) 
the same 1993 data sets as ours.
The difference is probably because of our selection of the energy band.
In fact, if we instead use 1--10 keV for instance, 
$Z_2^2=30.2$ and $\Pch=5.6 \times 10^{-4}$ are derived.

Then,  in  figure~\ref{fig:G9395_PGs}c, which fringes other than the highest f$_{0}$
should be considered as period candidates?
An obvious selection criterion is the statistical significance,
as implied by the $Z_2^2$ values  given in table~\ref{tbl:fringe_param}.
In Appendix A3, we evaluated the posterior probability $Q_n$ 
($n=-4$ to +4) for f$_n$ to represent the true pulse period,
all relative to $Q_0 \equiv 1$,
and quote the results in table~\ref{tbl:fringe_param}.
Because the period could still change to some extent
by the demodulation process to be conducted later,
at this stage we crudely exclude only those with $Q_n <0.01$, 
and conservatively retain four fringes from f$_{-1}$ to f$_{2}$.

As another criterion,
the black curve in figure~\ref{fig:G9395_PGs}c represents
a PG from G93pA +G93f (equivalently, G93T minus G93pB),
computed in the same way.
All the periods here refer to the epoch of the 1st photon in the G93pA data.
Although the values of $Z_2^2$ (right ordinate) are relatively low,
we still observe the fringe pattern,
of which the pitch, $\approx 60~\mu$s,  
is shorter by $\sim 10\%$ than in G93T.
This is because the effective data span of G93T is $\approx10$ d
as mainly determined by G93pB and G93f,
whereas that of G93pA+G93f is by $\sim 1$ d longer.
Thus, a ``vernier'' effect is created,
and $\delta P$ in table~\ref{tbl:fringe_param} gives the period difference 
between the corresponding red and black peaks. 
Requiring $|\delta P| \lesssim 16$ s
($\lesssim1/4$ of the fringe pitch),
we are left with the same four fringes
from f$_{-1}$ to f$_{2}$.

Since the above two criteria give consistent selections,
we tentatively  regard f$_{-1}$, f$_{0}$,  f$_{1}$, and f$_{2}$
as the pulse-period candidates as of 1993 October 9.
In table~\ref{tbl:results}, the G93T period and its uncertainty 
is hence defined so as to cover them.
The result is  consistent with those by KEA98 
(green arrows in figure~\ref{fig:G9395_PGs}c).

\begin{table}
\caption{Fringe parameters in the 2.8--12 keV PG from G93T.$^{*}$}
\label{tbl:fringe_param}
\begin{center}
\begin{tabular}{lccccr}
\hline 
Fringe  & $P_n^{ \dagger}$ &  $Z_2^2$  & $Q_n ^\ddagger$   &   $\delta P^\$$\\
num.& (s)&                                     \\
\hline \hline 
f$_{-4}$ &7.468\,209 (9) & 20.17 &$2.7 \times 10^{-4}$ &  $-36$   \\
f$_{-3}$ &7.468\,275 (8) & 25.96 & 0.0049&$-30$   \\       
f$_{-2}$ &7.468\,341 (8) & 24.92 &0.0029  &$-23$ \\
f$_{-1}$ &7.468\,412 (7) & 29.81 &0.034&$-10 $  \\
f$_{0}$  &7.468\,479 (7)& 36.58& (1) & $-3$\\
f$_{1}$  &7.468\,542 (7)& 34.54 & 0.36&+3  \\
f$_{2}$  &7.468\,607 (9)& 28.37 & 0.017&+9  \\
f$_{3}$  &7.468\,673(11) & 22.62 &$9.3 \times 10^{-4}$&+17 \\
f$_{4}$  &7.468\,737(13) & 16.55 &$4.5 \times 10^{-5}$ &+21 \\
\hline 
\end{tabular}
\begin{footnotesize}
\begin{itemize}
\setlength{\itemsep}{0mm}
\item[$^{*}$] Referring to the red PG in figure~\ref{fig:G9395_PGs}c.
\item[$^\dagger$] Defined at the start of the G93pA data stream.
\item[$^\ddagger$] Posterior robabilities, relative to f$_{0}$,
 for f$_{n}$ to be the true pulse period,
 calculated with equation~(\ref{eq:AppA4}) in Appendix A3.
\item[$^\S$] Period difference (in $\mu$s) of the red fringe peak,
relative to the corresponding peak in the black PG. 
\end{itemize}
\end{footnotesize}
\end{center}
\end{table}

\vspace*{-1mm}
\subsection{Demodulation analysis}
\label{subsec:ana_DeMD}

\subsubsection{Formalism}
\label{subsubsec:formalism}
Now that the pulsation has been reconfirmed in all data sets,
we proceed to the {\it demodulation analysis},
searching the data for the  PPM effects
(\cite{Makishima14};  2016; 2019).
For this purpose, the arrival time of each photon
(including background)
is changed from $t$ to $t-\delta t$ using 
\begin{equation}
\delta t = A \sin(2\pi t/T - \psi)~,
\label{eq:demodulation}
\end{equation}
where $T$, $A$, and $\psi$ denote
the period, amplitude, and initial phase of the assumed PPM, respectively.
Then, $T$ is varied from 7 ks to 100 ks, 
with a step of 0.1 ks to  0.5 ks.
At each $T$, we maximize the pulse significance,
by scanning $P$ over a $\pm 0.2$ ms interval (with $20~\mu$s step)
centered on the peak value in  figure \ref{fig:G9395_PGs},
$\psi$ over $0^\circ$ to $360^\circ$  with a step of $5^\circ$,
and $A$ from 0 to 1.5 s with 0.1 s step.
This range of $A$ is chosen
because $A$ is typically $\sim P/10$, 
and $\sim P/4$ at most \citep{Makishima16}.
The pulse significance is evaluated with $Z_4^2$,
where  $m=4$ is selected
because the demodulated profiles of magnetar pulses  
usually exhibit three to four peaks  (\cite{Makishima21a}).
The PPM is to be found only in the HXC signals,
and contamination of the SXC hampers its detection
more severely than the pulse detection.
We hence change the energy interval from 2.8--12 keV to 3--12 keV,
to more securely eliminate the SXC contribution
which could be significant up to $\sim 3$ keV.

\subsubsection{Results from individual data sets}
\label{subsubsec:results_individual}
\begin{figure}
\centerline{
\includegraphics[width=8.3cm]{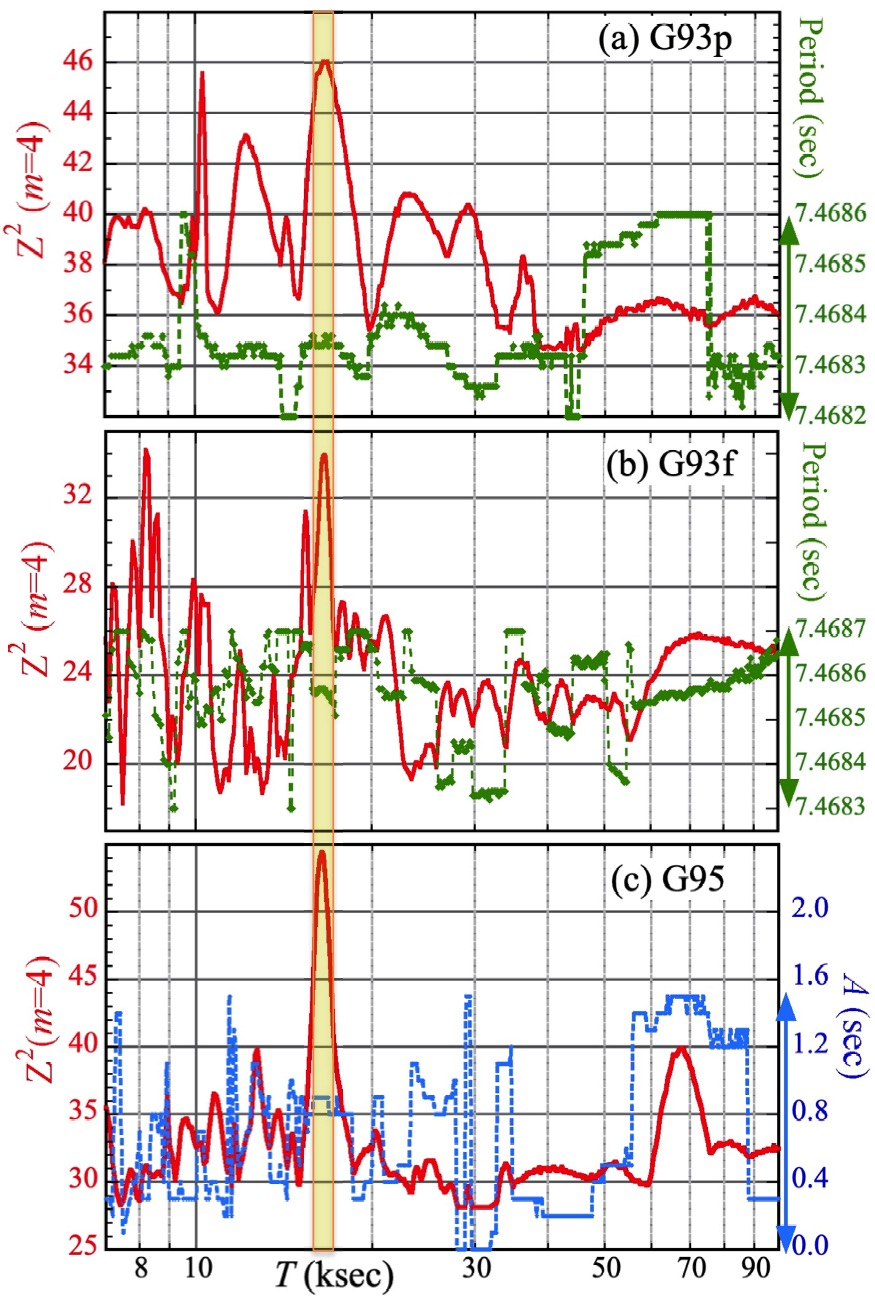}
}
\caption{DeMDs of \sgr\ in 3--12 keV (in red),
from (a) G93p, (b) G93f, and (c) G95,
where the highest $\zz$ value (left ordinate) is shown in red
as a function of the assumed modulation period $T$.
In (a) and (b), the best pulse period (right ordinate)
is given in green,
whereas in (c), the best modulation amplitude $A$  in blue
A vertical yellow stripe indicates a preferred common range of $T$.
 }
\label{fig:G9395_DeMD_indiv}
\end{figure}

By applying the demodulation analysis to the 3--12 keV data from G93p,
we obtained figure~\ref{fig:G9395_DeMD_indiv}a;
such a graph is called a {\em demodulation diagram} (DeMD),
which presents, against $T$,
the maximum $Z_4^2$  found at each $T$
when $A$, $\psi$, and $P$ are optimized.
A dominant peak has appeared at $T=16-17$ ks, 
and a sharp secondary one at $T\approx 10.5$ ks.
Table~\ref{tbl:results} summarizes the parameters of the former peak;
the errors (68\% confidence) associated with each quantity 
are determined as the points
where $\zz$ falls by 4.72  from  the peak value,
as explained in Appendix A2 and \citet{Makishima21a}.

The 16--17 ks peak is reconfirmed in the 3--12 keV DeMD from G93f,
shown in figure~\ref{fig:G9395_DeMD_indiv}b,
where $T$ is refined to $\approx 16.5$ ks (table~\ref{tbl:results}).
A similarly strong peak, observed at $T\approx 8.2$ ks, 
 could be a second harmonic
(half in the period and twice in the frequency).
As shown in green in figure~\ref{fig:G9395_DeMD_indiv}a 
and figure~\ref{fig:G9395_DeMD_indiv}b,
the pulse period depends to some extent on $T$.
Hereafter, the value of $P$ specified by the DeMD peak is denoted as $P_0$,
and regarded as the best-estimated pulse period for each data set
(except in G93T where the fringe ambiguity still remains).

The  PPM, suggested by the two 1993 data sets, 
becomes compelling in figure~\ref{fig:G9395_DeMD_indiv}c,
the DeMD from G95.
The  peak  indeed sticks out at $T \approx 16.4$ ks, 
and reaches $\zz =54.4$; 
this is comparable to the highest DeMD peaks
that were observed in the previous three magnetars \citep{Makishima21b}.
Thus, the  modulation period of $T  \approx 16.4$ ks $\equiv T_{16.4}$
is consistently preferred by the three data sets.
They also  agree  on $A=0.6-1.0 $ s,
whereas their differences in $\psi$ do not matter,
because $\psi$ is determined solely
by the start timing of the observation.

Taking the G95 DeMD as a typical case,
the significance of the  PPM was evaluated
through a  {\it control} study described in \citet{Makishima21b}.
That is, we derive a distribution of $\zz$  from the same G95 data,
at $T$ shorter than $\Tpk$ but much longer than $P$,
and compare the derived distribution of $\zz$
with the value  at $\Tpk$.
As given in Appendix B, 
the probability for the $\Tpk$ peak to appear by chance,
anywhere in the 7--100 ks range,  
is  estimated as  $< 1\%$.
This value will further decrease
when combining the G93p and G93f results.
Although $\Tpk$ is close to three times the ASCA's orbital period,
this artifact would appear at a discrepant period, $16.81 \pm 0.05$ ks,
as judging from the data-gap recurrence in the data.
We hence conclude, at 99\% confidence,
 that \sgr\ exhibits a PPM  at  $T\approx 16.4$ ks.
Our 1st objective has been fulfilled,
and \sgr\ becomes the fourth magnetar exhibiting this behavior.

\subsubsection{Demodulated PGs}
\label{subsubsec:demodulated_PGs}

\begin{figure}
\includegraphics[width=7.7cm]{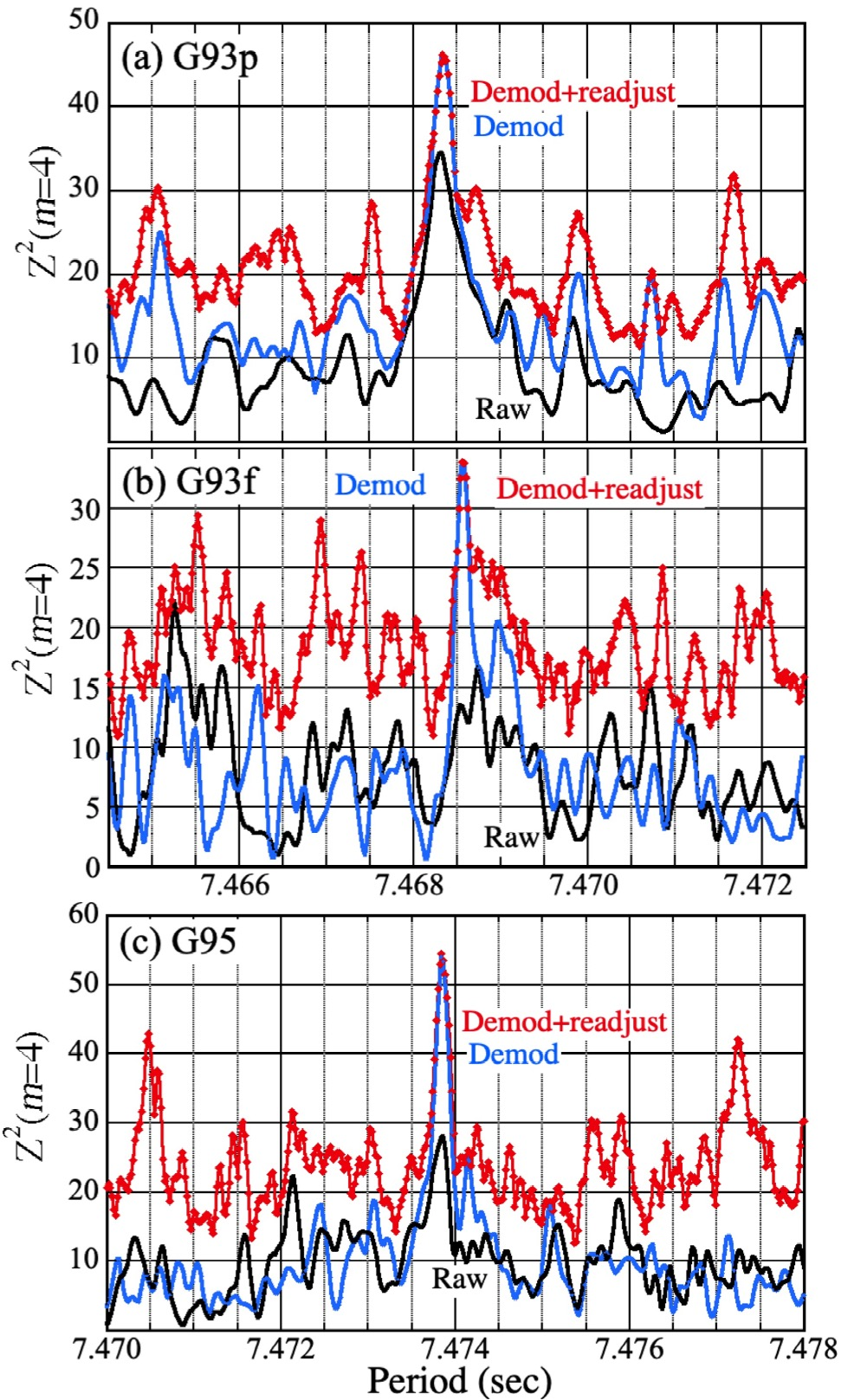}
\caption{
PGs in 3--12 keV, derived wth $m=4$ from the three data sets.
Black PGs are simple epoch-folding results,
whereas blue ones are those when the arrival times of all photons are
corrected by equation~(\ref{eq:demodulation}),
using a  parameter triplet $(T, A, \phi)$
given in table~\ref{tbl:results}.
The red PGs are derived by readjusting the parameter triplet at each $P$.
}
\label{fig:GIS9395_Pscan}
\vspace*{-3mm}
\end{figure}

Figure~\ref{fig:GIS9395_Pscan} provides
3--12 keV $Z_4^2$ PGs from the individual GIS data sets,
each computed under three different conditions.
Black shows a raw PG,
which differs from those in figure~\ref{fig:G9395_PGs}a
only in the energy range and the harmonic number.
The blue PG is produced through the PPM correction,
where we employ, at all $P$,
the triplet $(T, A, \psi)$ that is optimized at $P= P_0$
(table~\ref{tbl:results}).
Finally, the red PGs are those
in which the triplet is readjusted at each $P$;
we let $A$ and $\psi$ vary
in the same manner as in  figure~\ref{fig:G9395_DeMD_indiv},
whereas $T$ over its common uncertain range,
16.0--17.0 ks (table~\ref{tbl:results}) with a step of 0.1 ks.
These results afford the following inferences.

\begin{enumerate}
\item
The blue and red PGs agree well at $P\sim P_0$,
where the triplet $(T, A, \psi)$ converges
to that given in table~\ref{tbl:results}.
\item Except near $P_0$, the red PG is higher than the black and blue ones
by $\delta \zz \sim 10$,
due to an enlarged parameter space.

\item
The demodulation enhances the pulse significance
by $\delta \zz\approx 12$, 18, and  27,
in G93p, G93f, and G95, respectively.
Except in G93p, these exceed  the statistical increment in 2 above.
\item
The red  PGs (patricularly in c)  exhibit a pair of sub peaks,
spaced symmetrically from $P_0$ 
by $\delta P = \pm 3.4$ ms.
Since $\Tpk \delta P \approx P_0^2$ holds,
the sub peaks arise
when the interval $\Tpk$ contains $N\pm 1$ pulse cycles,
with $N \equiv \Tpk/P_0$.
This provides additional evidence
for the presence of double periodicity in the system, at $P$ and $T$.
The shorter sub peak can in fact be identified with $P_{\rm rot}$.
\item
The demodulated PG from G93f gives $P_0=7.46856(7)$ s  (table~\ref{tbl:results}),
which translates to $P_0=7.46849(8)$ s at the start of G93pA.
This error range accommodates  the fringes f$_{0}$ and f$_{\pm1}$,
even though the $P_0$ value from G93p (table~\ref{tbl:results})
favors somewhat shorter fringes, e.g., f$_{-3}$ through f$_{0}$.
We retain our conservative selection made in subsection~\ref{subsec:ana_PG},
to regard f$_{-1}$ through f$_{2}$ as the pulse-period candidates.
\end{enumerate}

\subsubsection{Behavior of the softest signals }
\label{subsec:ana_ASCA_softest}

Now that our 1st objective was fulfilled,
we move to the second aim, i.e., examinations
whether the PPM is associated with the HXC.
This is already accomplished partially,
because the PPM was detected from the three data sets
in the 3--12 keV interval.
We however need to confirm its absence at lower energies.

As a simple examination,
the lower bound energy  $E_{\rm L}$ for the DeMD calculation
was  gradually decreased from 3.0 keV,
with the upper boundary fixed at 12 keV.
Then, in the G93p data, 
the DeMD peak at $\Tpk$  became higher
until $E_{\rm L}$ hits $ \approx 2.55$ keV,
beyond which the peak  started diminishing.
In G93f and G95, this threshold was at $\approx 2.7$ keV.
The inclusion of photons below these thresholds 
suppresses the PPM;
the 3-12 keV energy range
which we have selected is considered appropriate.

\begin{figure}
\includegraphics[width=8.4cm]{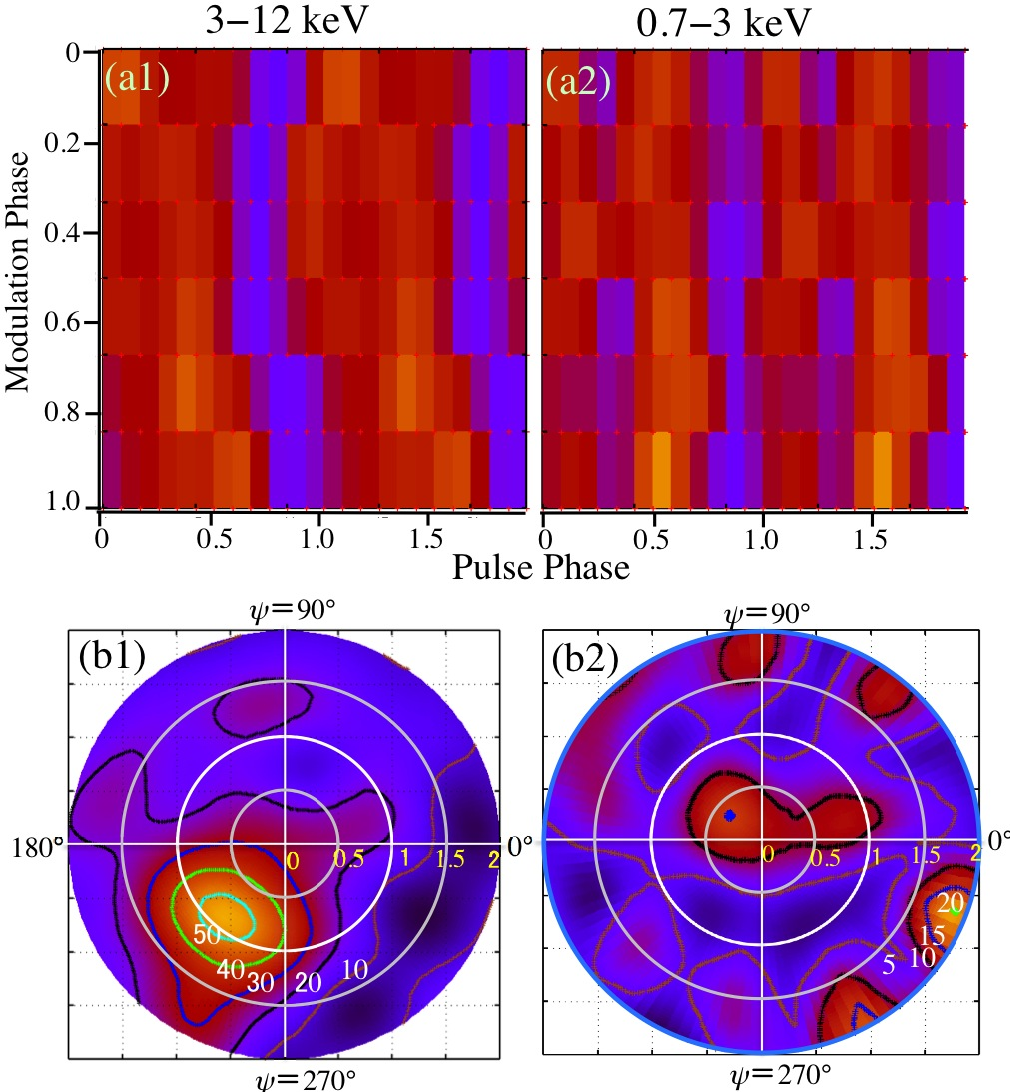}
\caption{
Top two panels are  color maps 
of the 3--12 keV (panel a1) and 0.7--3 keV (panel a2) photons from G95,
accumulated in two dimensions,
where the abscissa is the pulse phase (2 cycles) 
and the ordinate is the modulation phase.
Bottom two panes show,
on the polar coordinate  $(A, \psi)$,
the maximum $\zz$ achieved via the demodulation 
with equation~(\ref{eq:demodulation}). 
Panels (b1) and (b2) are for the 3--12 keV and 0.7--3 keV photons from G95,
respectively.
Contours in (b1) are from 10 to 50,
whereas those in (b2) are from 5 to 20.
}
\vspace*{-2mm}
\label{fig:G9395_2dplots}
\end{figure}

As a second attempt, we sort photons onto 
a pulse-phase vs.  modulation-phase plane,
and color-code the number of photons.
Then, the 3--12 keV data from G95 yielded
figure~\ref{fig:G9395_2dplots} (a1).
The pulse ridges, running vertically,
wiggle through the modulation phase,
visualizing the PPM effect.
The polar plot in figure~\ref{fig:G9395_2dplots} (b1) shows, 
as a function of  $A$ and  $\psi$,
the value of $\zz$ achieved in the demodulation
that employs $P \approx P_0$ and $T \approx \Tpk$.
The highest $\zz$ is found at $(A, \phi) = (0.9~{\rm s}, 235^\circ)$,
which is significantly  displaced from the coordinate origin.
In contrast, the 0.7--3 keV photons yield 
panels (a2) and (b2) of figurer~\ref{fig:G9395_2dplots}.
The pulse ridges are again clearly present, 
but they run straight.
The highest $\zz$ is hence obtained at $A \sim 0$
as in figure~\ref{fig:G9395_2dplots} (b2).
(The higher peak at $A \approx 1.9$ s should be ignored,
because we employ a limit of  $A \leq 1.5$ s.)
In a word, the  PPM at $\Tpk$ is present  in 3--12 keV of the G95 data,
but undetectable in 0.7--3 keV.

\begin{figure}
\begin{center}
\includegraphics[width=6.2cm]{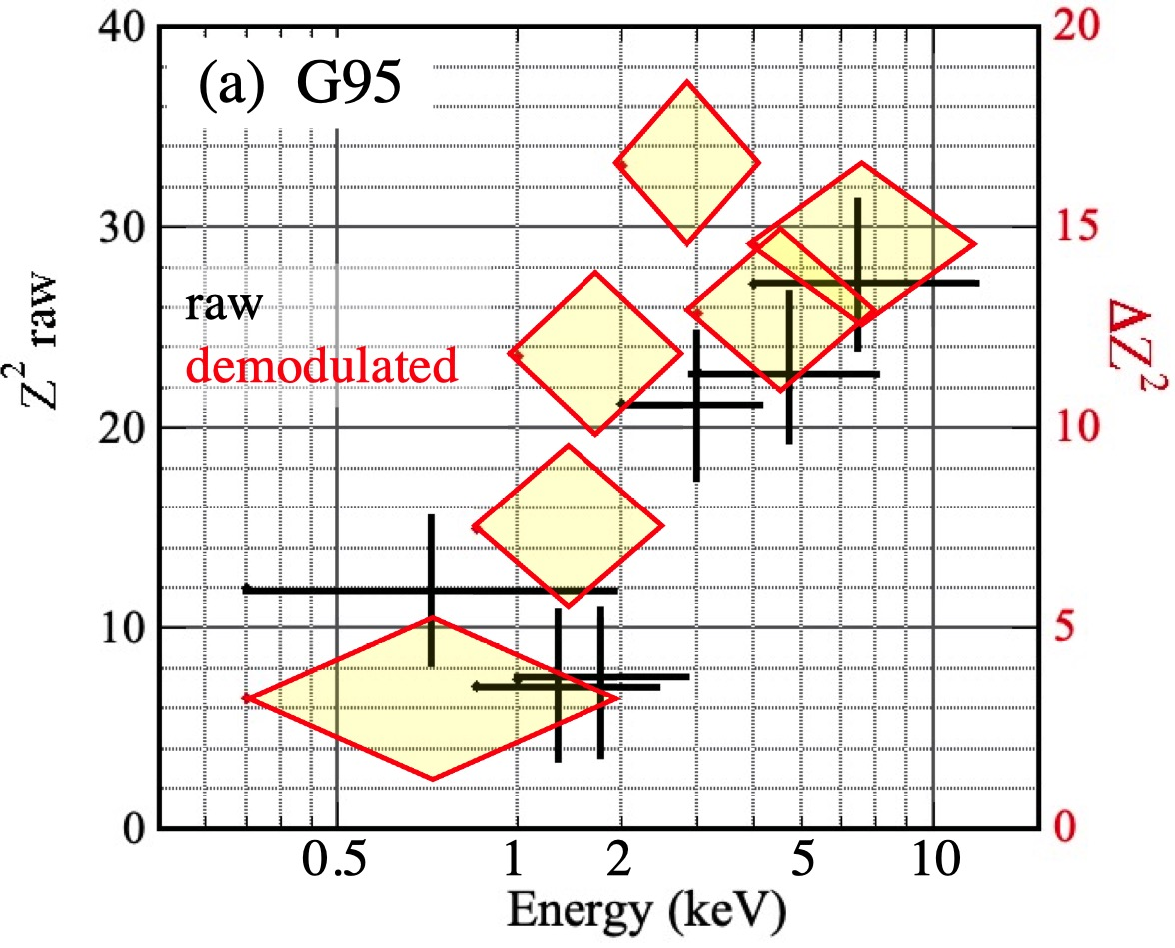}
\includegraphics[width=7.7cm]{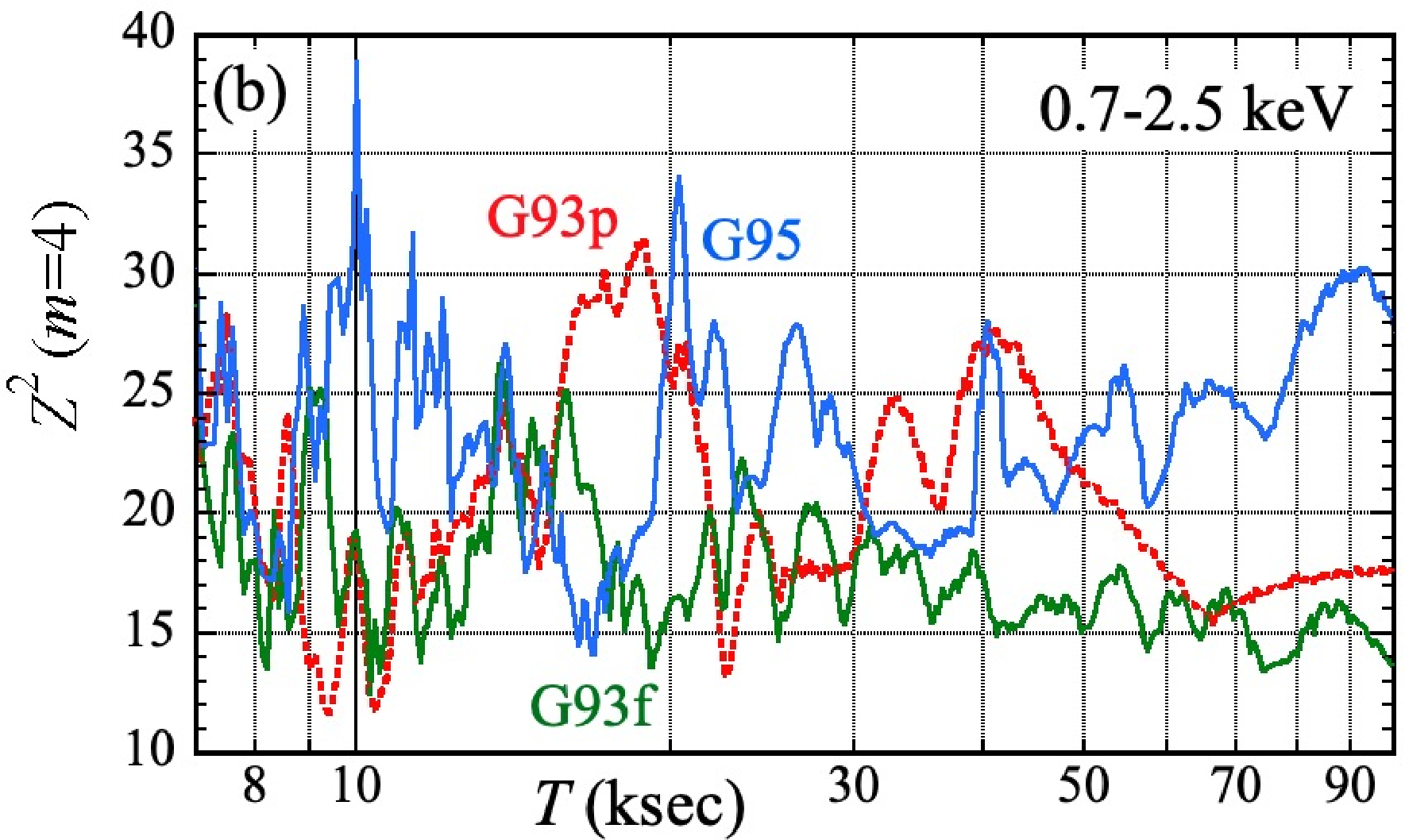}
\end{center}
\caption{(a) Pulse significance in G95  as a function of energy.
Black crosses show the $\zz$ values
(left ordinate)  at $P_0$ before demodulation,
whereas red diamonds represent the increments $\Delta \zz$ 
(right ordinate) achieved via demodulation,
of which the condition is detailed in text.
(b) DeMDs in 0.7--2.5 keV, from G93p (red),
G93f (green), and G95 (blue),
calculated in the same way as figure~\ref{fig:G9395_DeMD_indiv}.
}
\label{fig:G9395_SoftBandInfo}
\end{figure}

For a further confirmation, we calculated G95 DeMDs,
in seven partially-overlapping energy intervals 
with about the same photon numbers.
The result is given in figure~\ref{fig:G9395_SoftBandInfo}a,
where  black crosses show $\zz$  at $P\approx P_0$
before the demodulation,
whereas red diamonds give the increment $\Delta \zz$
attained  by the demodulation.
In obtaining the latter, we varied $P$ over $P_0 \pm 60~\mu$s,
$T$  between 16.0 and 17.0 ks, $A$ bewteen 0 to 1.5 s, 
and $\psi$ between $0^\circ$ to $360^\circ$.
Even though the pulse significance increases toward higher energies
already before demodulation,
$\Delta \zz$ rises more sharply at $\sim 2$ keV.

To examine the behavior of the softest signals
over a wider range of $T$,
we produced  0.7--2.5 keV DeMDs
from the individual data sets,
and show them together  in figure~\ref{fig:G9395_SoftBandInfo}b.
In the G93f and G95 results, the $\Tpk$  peak is no longer visible.
Although the G93p DeMD  exhibits a broad  hump over $T=16-19$ ks,
it  disappears when the upper energy bound is lowered from 2.5 keV to 2.1 keV.
The absence  of the  PPM in 0.5--2.5 keV is considered intrinsic,
rather than due to  insufficient statistics.

Through these multiple examinations,
we consistently found 
that the  PPM in \sgr\ is present in energies above $\sim 3$ keV,
whereas it is absent or much suppressed below $\sim 2.5$ keV.
This means an affirmative answer to our second aim.

\subsection{Demodulation analysis of the joint G93T data}
\label{subsec:ana_G93T}

\subsubsection{Procedure}

Our remaining task is to apply the demodulation analysis 
to the  merged G93T data, 
hoping to coherently phase-connect the 16.4 ks PPM
across the 11-d time span
which is dominated by the 8.8-d data gap.
If the modulation period $T$ is coherent,
it will also exhibit an interference structure,
like in figure~\ref{fig:G9395_PGs}c,
this time with a separation of $\Delta T \approx T^2/S \approx 0.27$ ks.
Thus, the search step in $T$ is selected to be 0.005 ks.
To cover the four surviving fringes from f$_{-1}$ to f$_{2}$ (figure~\ref{fig:G9395_PGs}c),
we scan $P$ from 7.46838 s to 7.46862 s with a step of  
$2~\mu$s  (the same as in figure~\ref{fig:G9395_PGs}c).
In contrast, $A$ and $\psi$ are treated in the same way as before,
because they are free from the interference.
Similarly, we retain $\dot P =0.8 \times 10^{-10}$ s s$^{-1}$,
which is now quite essential.
Based on the examination  in \S~\ref{subsec:ana_ASCA_softest},
the energy range is changed from 3--12 keV to 2.55--12 keV
to increase the signal statistics,
with the photon number increasing from 10971 to 12758.

\subsubsection{Results}
\label{subsubsec:demod_G93T}

\begin{figure}
\includegraphics[width=8.8cm]{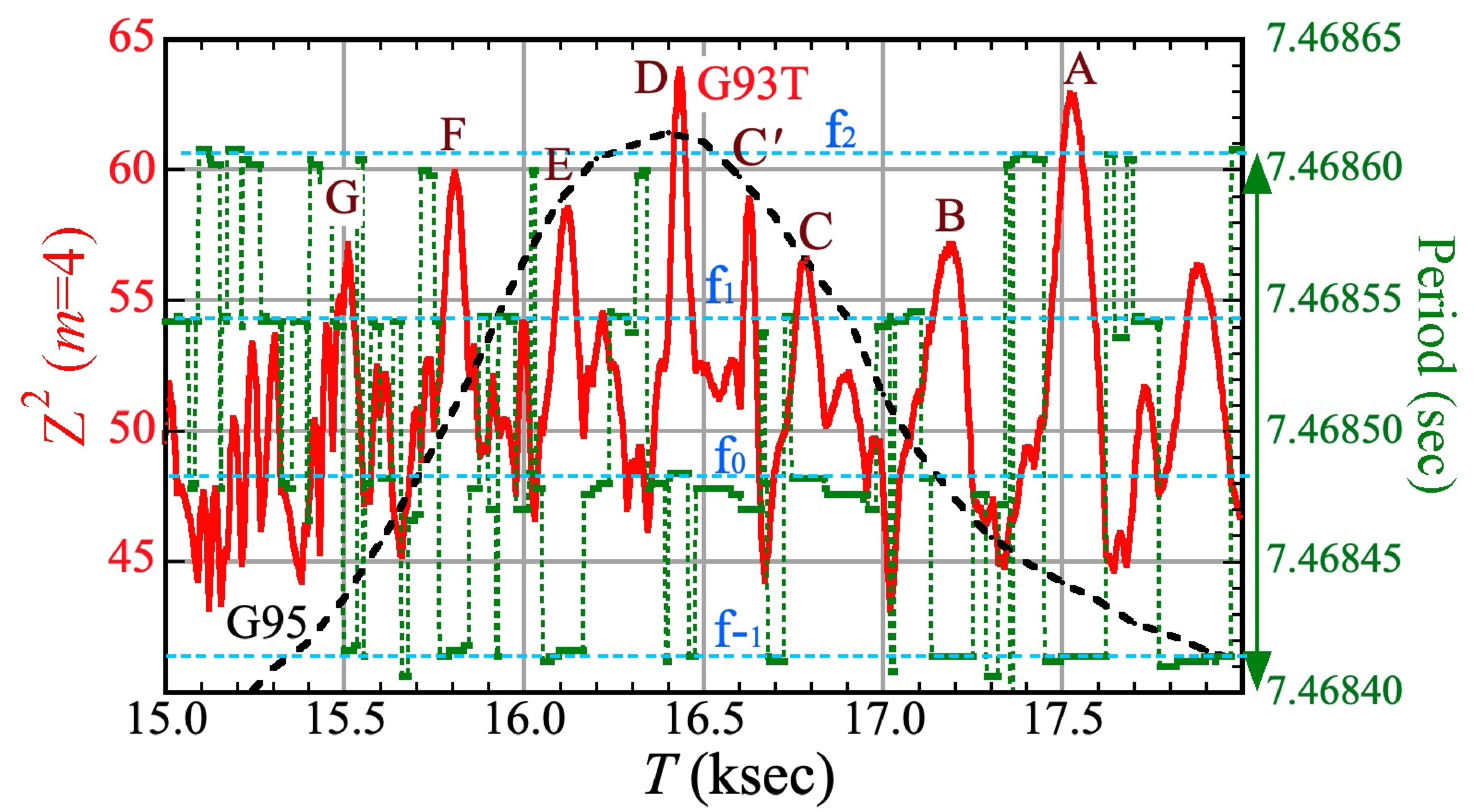}
\caption{A 2.55--12 keV DeMD from the combined G93T data (red),
compared with the 3--12 keV DeMD from G95 (black).
The latter is the same as in figure~\ref{fig:G9395_DeMD_indiv}c,
but is given a positive offset by 7 just for presentation.
Green data points show $P$ (right ordinate) 
associated with the G93T DeMD.
Dotted horizontal lines in cyan indicate the four surviving fringes
from figure~\ref{fig:G9395_PGs}c.
Peaks denoted as A through G are discussed in text.
}
\label{fig:G93T_DeMD}
\end{figure}

The 2.55--12 keV DeMD  thus derived from G93T is
presented in figure~\ref{fig:G93T_DeMD} in red.
As expected, we again observe clear interference peaks
denoted as A through G,
with an average pitch of $\sim 0.3$ ks.
Hereafter, we call these peaks ``$T$-fringes",
to distinguish them from those in $P$.
The structure is somewhat  more complex
than that in figure~\ref{fig:G9395_PGs}c,
and the pitch varies.
These properties are thought to arise 
because we now treat double periodicity, in $P$ and  in $T$.
The highest $T$-fringe, D ($\zz=63.86$), 
occurs at $T=16.435$  ks,
in a good agreement  with the peak of the 3--12 keV G95 DeMD,
which is superposed by a  dotted black curve with a positive offset by 7.
As shown in dotted horizontal lines in blue,
the $T$-fringe  D is associated with  $P=7.468484$ s (table~\ref{tbl:results}),
which coincides with f$_{0}$ in figure~\ref{fig:G9395_PGs}c.

Over $T=16-17$ ks in figure~\ref{fig:G93T_DeMD},
the envelope of the red curve, 
encompassing several  $T$-fringe peaks  such as C, C', D, and E,
is similar to the G95 DeMD.
Furthermore, in this interval,
the optimum $P$ mainly stays on f$_{0}$,
except several excursion episodes.
Outside this $T$ interval, in contrast, 
the G93T DeMD decreases much more slowly than that of G95,
and shows several ``outlier" peaks such as A, B, F, and G.
Since they all belong to the period fringes other than f$_{0}$,
they are thought to arise 
when pulse-phase mismatches across the long data gap,
due to  incorrect values of $P$,
are mitigated by changing $T$.

Although D is highest in figure~\ref{fig:G93T_DeMD},
the $T$-fringe A has nearly the same heigh; 
C' and F might also be considered.
This urges us to evaluate relative significances of these  $T$-fringe peaks.
The procedure is similar to that for the period  fringes 
made in section~\ref{subsec:ana_PG} and Appendix A3.
This time, however, we can use the G95 DeMD (black in figure~\ref{fig:G93T_DeMD})
as {\em prior} probability,
assuming $T$ to be the same between 1993 and 1995.
Then, the Bayes' theorem tells us how our knowledge on $T$ 
improves by adding the G93T DeMD.
We performed this evaluation  in Appendix A4, 
and derived the {\em posterior} probability
for  each $T$-fringe to represent the true value of $T$.
As a result, the peaks other than D were all found 
to have a probability which is $ \lesssim 3.4\%$ relative to D
(with C' the next most likely).
We can hence conclude, at 95\% confidence,
that D is the correct peak,
and  derive
\begin{equation}
T=16.435 \pm 0.024~{\rm ks}~~(T\mathrm{-fringe~D)}
\label{eq:T_93_best}
\end{equation}
as of the beginning of G93pA.

\begin{figure}
\includegraphics[width=8.2cm]{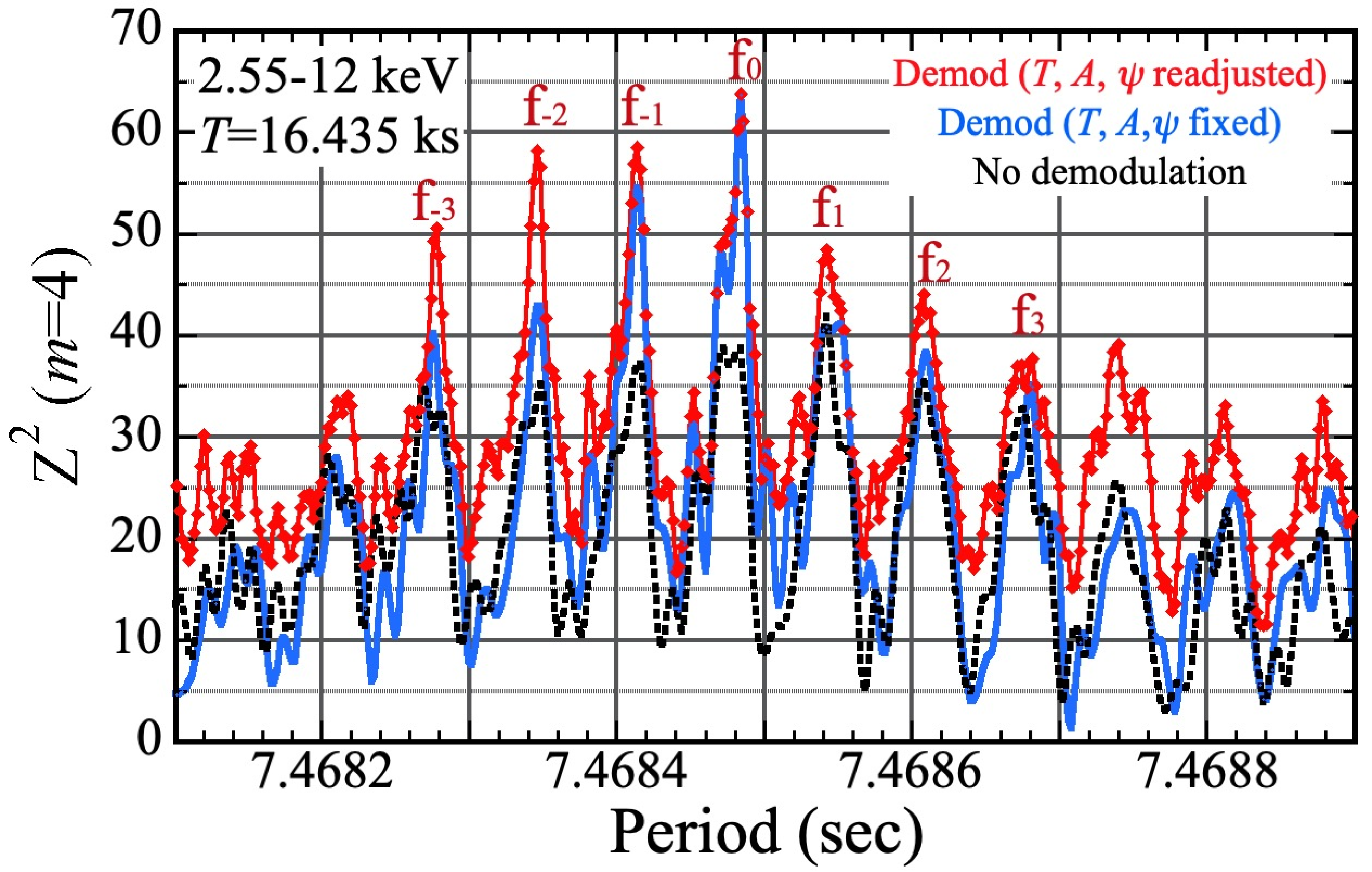}
\caption{PGs from the joint G93T data,
calculated in 2.55--12 keV with three different ways
as in figure~\ref{fig:GIS9395_Pscan}.
See text for details.
}
\label{fig:G93T_PG_3cases}
\end{figure}

\subsubsection{Demodulated PGs from G93T}

In figure~\ref{fig:G93T_DeMD},
the best $T$-fringe  D specifies f$_0$ as the best pulse period;
this combination can be expressed as (D, f$_0$).
However, it is not yet clear whether the other combinations
such as  (D, f$_{\pm 1}$) can be statistically ruled out.
To answer this issue,
figure~\ref{fig:G93T_PG_3cases} shows 
2.55-12 keV PGs ($m=4$) from G93T,
calculated under three different conditions
like in figure~\ref{fig:GIS9395_Pscan}.
The dotted black curve is the PG without demodulation,
which is similar to figure~\ref{fig:G9395_PGs}c
but differs in the energy range and $m$.
The blue one is a demodulated PG,
in which we  fix $T$ to equation~(\ref{eq:T_93_best}),
as well as $A=1.0$ s and $\psi=25^\circ$
as specified by the $T$-fringe D
of figure~\ref{fig:G93T_DeMD}.
When $A$ and $\psi$ are left to float at each $P$,
and $T$ is allowed to vary within $\pm 0.03$ ks of equation~(\ref{eq:T_93_best}),
the red PG is obtained.

The red PG in  figure~\ref{fig:G93T_PG_3cases}
gives $\zz=63.86$ for (D, f$_0$),
whereas $\zz=58.48$ for the next highest candidate, (D, f$_{-1}$).
According to the consideration in Appendix A4,
their height difference, $X=-\delta \zz=5.38$, implies 
that the relative probability for (D, f$_{-1}$) to be 
the true solution is $\exp(-5.38/2)=6.8\%$ of that of (D, f$_0$).
The candidates (D, f$_{n}$) with $n\ne 0$ are 
hence  excluded  at 90\% or higher confidence,
and the pulse period as of the start of G93pA is determined as
\begin{equation}
P=7.468\,484 (3) ~~{\rm s}~~({\rm fringe~f_{0}}).
\label{eq:P_93_best}
\end{equation}

If calculated in the same 2.55-12 keV interval as the G93T DeMD,
the G95 DeMD (thought not shown) has the maximum of $\zz=55.58$.
Because G93T has 1.4 times more photons than G95 (table~\ref{tbl:obs}),
and $\zz$ is proportional to the signal photon number
if the folded profile is similar,
we expect the peak D to reach $\zz =77.8$.
The actually observed peak D height,  $\zz=63.86$,
is $82\%$ of the prediction.
In addition, $\zz$ is proportional to the squared pulse fraction of the relevant periodicity,
where the pulse fraction is defined as $(H_{+}-H_{-})/(H_{+}+H_{-})$,
using the peak height $H_{+}$ and the bottom height $H_{-}$ of a folded pulse profile.
Accordingly, we infer that the PPM has been phase-connected across the 11-d time span
with a good phase coherence of $\sqrt{0.82}=0.91$.

In this way, the demodulation of the merged G93T data
in 2.55--12 keV enabled us 
to confirm the coherence  of the PPM cycle across the 11-d interval,
and measure its  period $T$ with an unprecedented accuracy of 0.15\%.
These are thanks to the particular observation patter in 1993,
and  to the shortest $T$ among 
the four magnetars exhibiting the PPM.
Thus, our 3rd objective has also been fulfilled.
At the same time, the demodulation analysis successfully resolved
the period ambiguity which was identified in figure~\ref{fig:G9395_PGs}c.

\begin{figure}
\includegraphics[width=6.5cm]{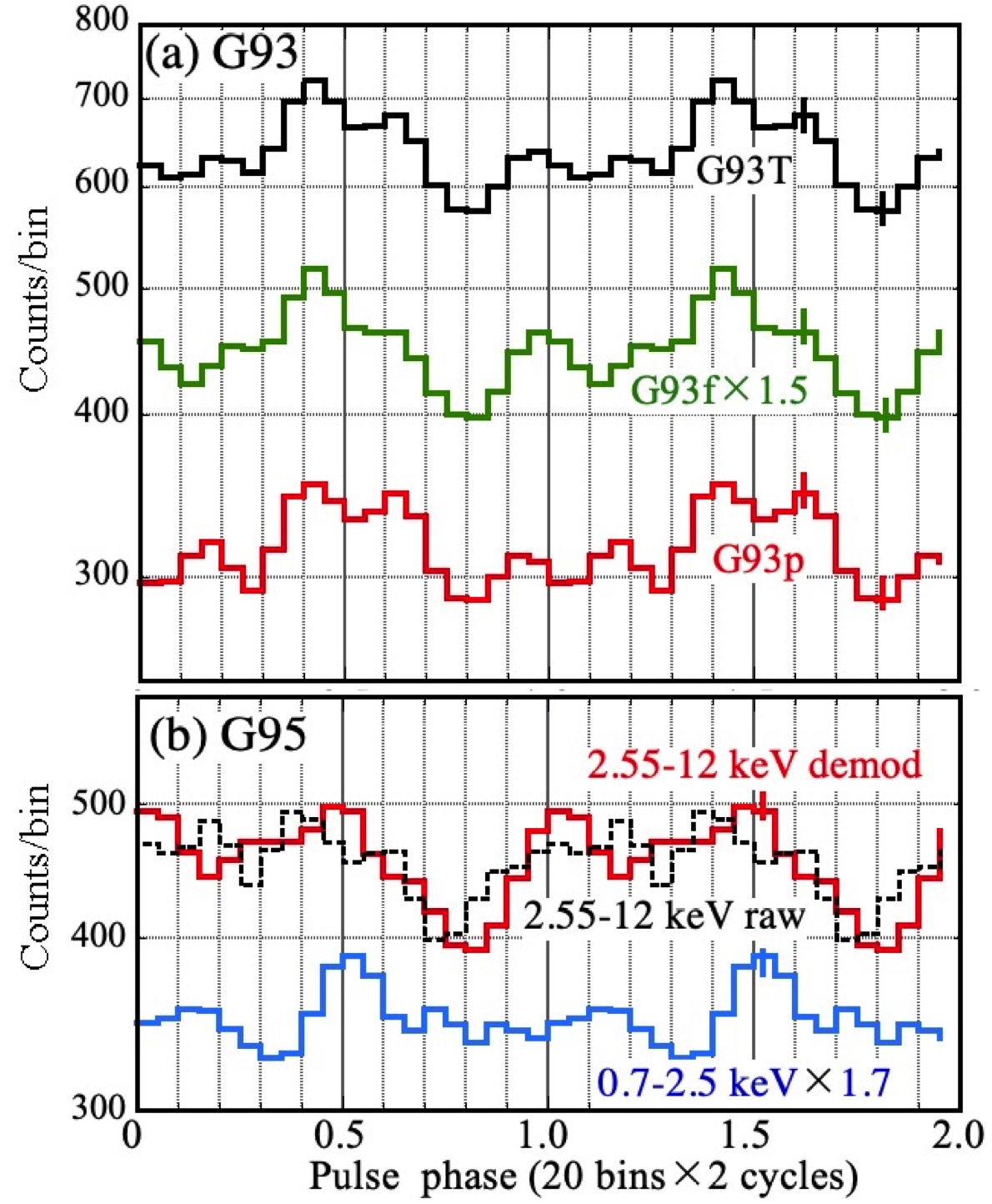}
\caption{Background-inclusive pulse profiles of \sgr\ 
from the GIS data sets, shown for two pulse cycles
after applying the running average (see text).
(a) Results in 2.55--12 keV  from G93p (red), 
G93f (green) multiplied by 1.5, and G93T(black).
They are all folded using a common solution (see text).
(b) The 2.55-12 keV profiles from G95,
before (black) and after (red) the demodulation.
The phase origin is manually adjusted to that of (a).
A 0.7--2.5 keV profile, folded without demodulation,
is presented in blue.
}
\label{fig:PulseProfiles}
\end{figure}

\subsubsection{The spin-down rate }
From the 1993 and 1995  period measurements  (table~\ref{tbl:results})
across a span of $\approx 2$ yrs,
the average spin-down rate is derived as 
$0.84\times 10^{-10}$ s s$^{-1}$,
in agreement with KEA98.
In addition, the long time span of G93T may allows us to estimate 
the instantaneous $\dot P$ in 1993 September.
By changing it from 0 to $1.5\times 10^{-10}$ s s$^{-1}$,
we studied how $\zz$ of the 
demodulated 2.55-12 keV pulses from G93T varies.
As a result, a loose constraint as
$\dot P = (0.77 \pm 0.63) \times 10^{-10}$ s$^{-1}$
(68\% confidence)
was obtained. From these two estimates, 
our assumption of $\dot P = 0.8\times 10^{-10}$ s$^{-1}$,
which has so far been employed, is confirmed to be appropriate.

\subsection{Pulse profiles}
\label{subsec:ana_Pr}
In figure~\ref{fig:PulseProfiles}, we compile background-inclusive 
folded pulse profiles obtained from the GIS data. 
They are smoothed with a running average as described in \citet{Makishima21a},
so the error associated with each bin is 0.61 times the Poissonian value.
The ordinate is logarithmic,
to enable a direct comparison of the pulsed fraction among the profiles.
Although the 2.55--12 keV photons are used here,
the results are very similar even if using the 3--12 keV interval.

Figure~\ref{fig:PulseProfiles}a compares 
2.55--12 keV profiles from G93p, G93f, and G93T.
They are all obtained using a common solution specified  by
equation~(\ref{eq:P_93_best}) and equation~(\ref{eq:T_93_best}), 
together with $A$ and $\psi$ for G93T in table~\ref{tbl:results},
and shown against a common pulse phase.
The three profiles are reasonably similar,
with the deep minimum at a pulse phase 0.8,
and nearly the same pulse fraction.
These demonstrate that the pulse phase has been
successfully connected through the 11-d interval.
The G93p result comprises equally-spaced four peaks,
whereas two of them become dominant in the G93f result.

Figure~\ref{fig:PulseProfiles}b compares the 2.55--12 keV
profiles from G95, before (dashed black) and after (red) the demodulation.
The demodulated profile is approximately double peaked,
which is closer to the G93f profile than to that in G93p.
This kind of mild profile variations  are
often seen in magnetars \citep{Makishima21b}.
Also shown is the 0.7--2.5 keV raw profile,
representing the SXC pulses,
to be compared with panel (a2) of figure~\ref{fig:G9395_2dplots}.

\begin{table*}
\caption{PPM parameters of the four magnetars.}
\begin{small}
\begin{center}
\label{tbl:PPM_param}
\begin{tabular}{lcccccccccc}
\hline 
Source  & Data $^*$& Energy & $P$ & $T$ (ks)  
      & $P/T $ & $B_{\rm t}^\dagger$& $B_{\rm d}^\ddagger$& $B_{\rm t}/B_{\rm d}$ &$\tau_{\rm c}^\S$ & Ref.$^\|$\\
           &                & (keV) & (s) & (ks) & $(10^{-4})$  &  &  &    & (kyr)\\
\hline \hline 
4U 0142+61  & Suzaku (07)& 15--40 & 8.6888& \multicolumn{5}{c}{PPM not detected}& &[1,2]\\ 
                      & Suzaku (09)& 15--40 & 8.6889& $55 \pm 4 $ &1.6    &  1.3 &1.3 & 100 &68 & [2]\\                         
                      & Suzaku (13)&15--40& 8.6891 & $54 \pm 3$ 
                      & \multicolumn{5}{c}{--- the same as above ---}  &[3]\\                          
                      & NuSTAR (14)&10--70& 8.6892 &$54.8 \pm 5.3$ 
                      & \multicolumn{5}{c}{--- the same as above ---}  & [3]\\ 
\hline 
SGR 1900+14 &  Suzaku (09) & 12--50 & 5.210 & $41.2 \pm 1.2$ & 1.3  & 1.1 & 7.0 &16 & 0.90 &[4] \\
                      & NuSTAR (16) &  6--20 & 5.227  &  $40.5 \pm 0.8$ 
                      &\multicolumn{5}{c}{--- the same as above ---}   &[4] \\
\hline 
\oneE            & Suzku  (09)&10-30  &2.072 & $36.0^{+4.5}_{-2.5}$ &0.58 & 0.76 &3.2 & 24 &0.69 & [5]   \\
                    &   NuSTAR (16) &  8--25 &2.087 & $36.0 \pm 2.3 $  
                     &\multicolumn{5}{c}{--- the same as above ---}  & [6] \\
\hline 
SGR 1806-20 & ASCA (93)   & 2.55--12 & 7.4685 & $16.435 (24)$ & 4.5 &2.1&20 &11 &0.24 & This work \\
                      & ASCA (95)    & 3--12     & 7.4739 & $16.4 \pm 0.4$ 
                      &\multicolumn{5}{c}{--- the same as above ---}   &This work \\
\hline
\end{tabular}
\end{center}
\end{small}
\begin{footnotesize}
\begin{itemize}
\setlength{\itemsep}{0mm}
\item[$^{*}$] The utilized data, and the last two digits of the observation year in parenthesis.
The results from the Suzaku XIS are not shown.
\item[$^\dagger$] Toroidal magnetic fields in units of $10^{16}$ G, 
calculated with equation~(\ref{eq:Bt_epsilon})
assumig $\cos \alpha \sim 1$.
\item[$^\ddagger$] Surface dipole magnetic fields in $10^{14}$ G, 
calculated as $\propto (P \dot P)^{1/2}$,
taken from McGill Online Magnetar Catalog (http://www.physics.mcgill.ca/~pulsar/magnetar/main.html)
\item[$^\S$] Characteristic age $\tau_{\rm c} \equiv P/2 \dot {P}$.
\item[$^\|$] References: [1] \citet{Enoto11}; [2] \citet{Makishima14} ; [3] \citet{Makishima19}; 
 [4] \citet{Makishima21b};   [5] \citet{Makishima16}; [6] \citet{Makishima21a}; 
\end{itemize}
\end{footnotesize}
\end{table*}

\section{Discussion}
\label{sec:discussion}

\subsection{Summary of the results}
\label{subsec:summary_results}

We analyzed the 0.7--12 keV X-ray data of \sgr\
acquired with the \ASCA\ GIS,
on three occasions  in 1993 October (G93pA, G93pB, and G93f),
and  once in 1995 October (G95).
The aims are threefold;
to search for the PPM  phenomenon,
to confirm its  association with the spectral HXC,
and to coherently connect the PPM phase
across the 11-d interval spanned 
by the merged 1993 data (G93T=G93p+G93f).

The periodogram analysis with $m=2$ (figure~\ref{fig:G9395_PGs}a),
using 2.8--3 keV photons from G93p, G93f,
and G95, individually yielded the pulse periods 
as in table~\ref{tbl:results}.
The combined data G93T revealed a clear
interference pattern (figure~\ref{fig:G9395_PGs}c),
and the probability for the highest fringe to appear by chance fluctuations
was estimated as $\Pch=2.8 \times 10^{-5}$
taking into account ``look-elsewhere" effects.
These results reconfirm the pioneering work by KEA98
who analyzed the same GIS data.

To these data sets,  we  next applied the demodulation analysis 
using the 3--12 keV energy interval and $m=4$.
Then, the G95 data gave a highly significant ($\gtrsim 99\%$; Appendix B)  
DeMD peak  at $T \approx 16.4$ ks (figure~\ref{fig:G9395_DeMD_indiv}c).
The same effect, though weaker,
was confirmed in the 3--12 keV DeMDs from
G93p  (figure~\ref{fig:G9395_DeMD_indiv}a) 
and G93f (figure~\ref{fig:G9395_DeMD_indiv}b).
We hence conclude that \sgr\ is the fourth magnetar 
exhibiting the PPM phenomenon, with the shortest $T$.
The effect was  confirmed down to $\sim 2.5$ keV,
but absent in lower energies
(figure~\ref{fig:G9395_2dplots}b, figure~\ref{fig:G9395_SoftBandInfo}).
Thus, the 1st and second aims of the research have been accomplished.

Finally, we applied the demodulation analysis to the merged G93T data in 2.55--12 keV,
and successfully connected the PPM phase through  the entire G93T data,
which is 11-d long, or about 58 modulation cycles.
We thus confirmed the PPM to have high stability,
and determined $T$ with an unprecedented accuracy of 0.15\%.
The demodulation resolved the period ambiguity in figure~\ref{fig:G9395_PGs}c,
and yielded equation~(\ref{eq:P_93_best}) and equation~(\ref{eq:T_93_best}).
In short, we successfully achieved the third objective as well.

\vspace*{-2mm}
\subsection{Ubiquity of the phenomenon}
\label{subsec:ubiquity}

Among about 30 magnetars known to date,
\sgr\ is the prototypical object with extreme activity,
showing the highest $B_{\rm d}$ 
and the shortest characteristic age, $\tau_{\rm c} \equiv P/2 \dot{P}$.
Therefore, \sgr\  was a sort of  ``touchstone''
to assess whether  the PPM is generally seen among magnetars.
The present result has indeed given an affirmative answer to this anticipation.

Table~\ref{tbl:PPM_param} summarizes the basic PPM parameters
of the four magnetars (including \sgr)
that exhibit this phenomenon.
Regarding  G93T and G95 of \sgr\ as two independent data,
so far the 10 data sets listed in this table have been analyzed
for the PPM effect.
It  was detected successfully from all these data,
except the 2007 \Su\ observation of 4U~0142+64.
Therefore, the PPM must be ubiquitous among magnetars.
The sole exception may be explained 
if the PPM amplitude then happened to be very small, e.g., $A \lesssim 0.05P$, 
as actually found with \NuS\ in 2009 \citep{Makishima21a}
thanks to its higher sensitivity than the \Su\ HXD.

When studying a new phenomenon like the present subject,
we must carefully exclude instrumental artifacts.
In this respect, the present result is of high value,
because it means that the PPM  in the four magnetars 
has been confirmed by four instruments;
the HXD and the XIS onboard Suzaku, NuSTAR, and the ASCA GIS.
They respectively utilize non-imaging Silicon PIN diodes,
X-ray CCD devices, pixellated CdZnTe detectors, and imaging gas scintillation proportional counters,
with the latter three coupled with X-ray focusing telescopes.
The detections with these distinct types of X-ray detectors,
onboard three different spacecrafts, is thought to minimize 
the risk of the phenomenon being of some instrumental origin.

It is intriguing to examine
how the PPM  phenomenon in magnetars (including \sgr)
remained undetected until 2014.
Presumably, past timing studies of magnetars
mostly utilized SXC-dominated energies  (e.g., $\lesssim 10$ keV) 
which are usually free from the PPM disturbance.
Even when the utilized instrument (e.g., RXTE) covers 
HXC-dominated energy ranges,
usually the period found in the softer band was used 
to produce folded pulse profiles in harder X-rays,
without examining whether  the pulse is significant therein or not.
Obviously, the situation is different for \sgr\ in 
which the HXC dominates down to $\sim 3$ keV,
but probably the pulse detection in this particular object
was possible without demodulation,
as we have experienced (figure~\ref{fig:G9395_PGs}).
The PPM perturbation might be 
smeared out due to the short $T$.

For reference,
our first detection of the PPM phenomenon, 
from 4U~0142+61 \citep{Makishima14},
was enabled by good fortune.
The 15--40 keV \Su\ HXD data of this object, acquired in 2007,
allowed the pulse detection via a standard periodogram analysis \citep{Enoto11}.
In contrast,  from the HXD data acquired in 2009,
the same analysis in the same energy band 
failed to detect the pulsation,
in spite of very similar exposures and source intensities.
Presumably $A$ changed as mentioned above.
This discrepancy between the two observations
drove us to the PPM discovery.

\subsection{Astrophysical implications}
\label{subsec:implications}

\subsubsection{The free-precession interpretation}
\label{subsubsec:free_precession}

As describe in section \ref{sec:intro}, 
the PPM effect in magnetars has so far been interpreted 
as a manifestation of the free precession of an NS
that is axially deformed to $\epsilon \approx P/T \sim 10^{-4}$. 
In table~\ref{tbl:PPM_param}, the ratio $P /T = 4.5 \times10^{-4}$, 
which we find for \sgr, is higher than, 
but still of the same order as, those of the other three magnetars. 
Moreover, the modulation amplitude of \sgr, 
$A = 0.6-1.0$ s (table~\ref{tbl:results}) or $(0.08-0.13)P$, 
agrees with those of the others. 
Therefore, the free-precession interpretation 
will apply also to this prototypical magnetar.

By the successful phase connection through the G93T data spanning 1000 ks, 
the PPM in \sgr\  was confirmed to have good coherence 
for at least $\sim 60$ cycles. 
As a result, $T$ was determined with an accuracy of 0.15\%. 
These results mean large improvements over our past knowledge, 
where the reproducibility of  $T$ of each object was confirmed 
only to an accuracy of $\sim10\%$ (table 3). 
We can now regard the PPM as a stable periodic phenomenon, 
rather than some transient quasi-periodicity. 
Since such high stability is likely to arise via celestial mechanics, 
the free precession scenario is greatly reinforced.

\begin{table}
\caption{Classification of rotational motions of a rigid body.$^{*}$}
\label{tbl:symmetry}
\begin{tabular}{lcccccl}
\hline 
Case &Mode of      & \multicolumn{3}{c}{Condition$^\dagger$} & $T$ of & Pulse \vspace*{-1mm}\\
       & rotation     &  (i)  & (ii)  & (iii)  &  eq.(\ref{eq:slip_period}) & ~detection\\
\hline 
\hline 
A1 &Spherically  &yes &yes &yes & $\infty$& no \vspace*{-1mm}\\
A2 & symmetric  &yes & yes &no& $\infty$ &\footnotesize{ $P_{\rm rot}(=P_{\rm pr})$ }\vspace*{-1mm}\\
A3  &  rotator &yes &no& $\ddagger$ & $\infty$&\footnotesize{ $P_{\rm pr}(=P_{\rm rot})$ }\vspace*{1mm}\\
\hline 
B1 & ``Sleeping" &no &yes &yes & finite & no\vspace*{-1mm}\\
B2 &  rotator      &no&yes &no &finite & $P_{\rm rot}$\vspace*{1mm}\\
\hline 
C1 &Free  &no&no&yes & finite &$P_{\rm pr}$ \vspace*{-1mm}\\
C2&precession &no&no&no & finite &\footnotesize{$P_{\rm pr}\times $PPM} \vspace*{1mm}\\
\hline 
\end{tabular}
\begin{footnotesize}
\begin{itemize}
\item[$^*$] Non-axi-symmetric deformation is not considered.
\item[$^\dagger$]  See text for the meaning of the three conditions.
\item[$^\ddagger$] The pulse behavior does not depend on this condition.
\end{itemize}
\end{footnotesize}
\end{table}

The overall phenomenon may be understood
by considering the following three symmetry conditions,
which describe an axisymmetric rigid rotator  
emitting radiation \citep{Butikov06}.
\begin{itemize}
\item[(i)] The object is spherically symmetric; namely, $\epsilon=0$. 
\item[(ii)] The symmetry axis $\hat x_3$ is aligned with $\vec{L}$; namely, $\alpha=0$.
\item[(iii)] The emission is axially symmetric around $\hat x_3$.
\end{itemize}
Then, the pulse detectability can be classified as in table 4, 
according to whether each condition holds (denoted by ``yes") or is violated (denoted by ``no").
Among the total seven cases, A2 and A3 apply to ordinary spherical pulsars.
These two are in reality equivalent, 
because A3 reduces to A2  if $\hat x_3$ is redefined 
as a unit vector parallel to $\vec L$.

The remaining cases ($\epsilon \ne 0$)
can be best explained in terms of a spinning rugby ball,
which obviously violates the condition (i).
When the ball is spinning precisely around  $\hat x_3$ ($\alpha=0$),
and if its appearance is symmetric around $\hat x_3$,
we can never tell that the ball is spinning (case B1).
However, if the ball has a ``logo mark" pained on its side,
(iii) is violated, and we can detect the spin by the logo's recurrence.
This corresponds to B2, and the observed pulse period $P$ gives $\Prot$,
with $\Ppr$ undetectable.

If $\alpha \ne 0$ instead, the rugby ball wobbles, 
in such a way that its tips rotate around $\vec L$ with a period $\Ppr$,
which satisfies $\Ppr:\Prot \approx 5:3$.
If the ball has no logo [(iii) satisfied],
we can detect only $\Ppr$ via the tip recurrence.
This is the case C1, 
wherein the pulse period to be observe is $\Ppr$ rather than  $\Prot$ (secion~\ref{sec:intro}).
This condition accounts for the magnetar SXC,
which may be emitted isotropically from the magnetic poles (=around $\hat{x}_3$) .
Finally, the HXC of magnetars  corresponds to C2,
wherein the basic periodicity is still $\Ppr$,
but it is coupled with $\Prot$ to exhibit the PPM;
the amplitude $A$ depends on $\alpha$, 
the degree of emission asymmetry around $\hat{x}_3$,
and our viewing angle to the system
\citep{Makishima21a}.

Identifying  the observed $P$ of a magnetar with $\Ppr$
may contradict to the common belief to identify $P$ with $\Prot$.
However, $\Ppr$ and $\Prot$ of a magnetar differ by ony $\sim 10^{-4}$,
and both are directly proportional to $L$,
with their ratio kept strictly constant
(assuming $\epsilon$ and $\alpha$ to be constant).
Hence, $\Ppr$ and $\Prot$ increase together
due to the same mechanism, namely, 
a decrease in $L$ through, e.g., the magnetic dipole radiation.
If a rugby ball with $\alpha \ne 0$ spins down,
its wobbling inevitably gets slower, too.
Thus, the general framework of estimating $\Bd$ and $\tau_{\rm c}$,
from the observed  $P$ and $\dot P$, still remains intact.

When $L$ of the object 
is conserved and its  kinetic energy decreases,
the wobbling angle $\alpha$ approaces 0 if  the deformation is oblate ($\epsilon <0$),
whereas $|\alpha|$ increases if prolate ($\epsilon >0$).
At present, the observations are unable to tell whether 
the deformation of these magnetars is oblate or prolate.
Nevertheless, we believe that the deformation is prolate,
because $\alpha$ would then develop,
with no external perturbation, to a finite level,
even if it was initially very small.

\subsubsection{Toroidal magnetic fields of magnetars}
\label{subsubsec:Bt_of_magnetars}

Taking the free-precession scenario for granted,
we can think of three origins of the deformation;
centrifugal force, magnetic stress of $B_{\rm d}$,
and that of $B_{\rm t}$.
The deformation from the first two origins will be oblate,
and are too small to explain $|\epsilon| \sim 10^{-4}$.
In contrast, the toroidal field hidden inside the NS
will cause a prolate deformation.
Furthermore, if $B_{\rm t} \sim 10^{16}$ G,
the deformation will be large enough to explain the observation
as in equation~(\ref{eq:Bt_epsilon}).
Using this relation,
we  calculated $B_{\rm t}$ 
and give the results for the four magnetars
in table~\ref{tbl:PPM_param},
together with their $B_{\rm d}$ and $\tau_{\rm c} $.
Thus, the four objects are all inferred to harbor 
ultra-strong toroidal magnetic fields,
$B_{\rm t} \sim 1 \times 10^{16}$ G.

Importantly, table~\ref{tbl:PPM_param} comprises objects 
in all the three major subclasses of magnetars:
two Soft Gamma Repeaters (SGR 1900+14 and \sgr),
an Anomalous X-ray Pulsar (4U~0142+61), 
and one transient magnetar (\oneE).
When combined with the discussion in  subsection~\ref{subsec:ubiquity},
this suggests that essentially all magnetars harbor $B_{\rm t} \sim 10^{16}$ G,
even though the examples are still  limited at present.

The four objects in table~\ref{tbl:PPM_param} are
arranged in the descending order of their $\tau_{\rm c}$.
We observe a trend of the $B_{\rm t}/B_{\rm d}$ ratio
to increase toward older magnetars.
Therefore, the toroidal fields of magnetars 
possibly  last longer than their dipole fields
\citep{Makishima23}.
This evolution, if confirmed with a larger sample,
will naturally explain the presence of such NSs as
weak-field magnetars (e.g., \cite{Rea14}).
It is however not obvious 
whether the activity of magnetars is powered
by $B_{\rm d}$ which is weaker but supposedly more easily dissipated,
or $B_{\rm t}$ which is stronger but could be more difficult to utilize.
In any case, we need a larger sample.
In addition, a more quantitative study is necessary,
because $\tau_{\rm c}$ of magnetars, 
which is calculated assuming a constant $B_{\rm d}$,
is likely to  overestimate their true age \citep{Nakano15}.

\vspace*{-1mm}
\subsubsection{The two spectral components of magnetars}
\label{subsubsec:two_spectral_components}

Persistent X-ray spectra of magnetars are known 
to consist generally  of
the HXC and the SXC (section~\ref{sec:intro}; \cite{Enoto10}),
which are characterized respectively
by a blackbody-like spectral shape and a hard power-law form.
In 4U~0142+61, \oneE, and SRG~1900+14,
the PPM was found to disappear below $\sim 10$ keV,
$\sim 8$ keV, and $\sim 6$ keV, respectively,
which are  close to their HXC vs SXC cross-over energies.
Therefore, he PPM perturbation was suggested to affect
only the HXC pulses (e.g., \cite{Makishima16}).

In the present work, we utilized the very bright HXC of \sgr,
to confirm that the PPM is present down to $\sim 2.55$ keV,
again close to the cross-over energy of the two components.
The exclusive association of the PPM with the HXC
has been reinforced.
The presence/absence of the PPM at a given energy is 
likely to be determined by the relative dominance 
of the two components at that energy,
rather than the absolute value of the energy itself.

As mentioned with respect to table~\ref{tbl:symmetry},
a likely  origin of this HXC vs  SXC difference 
is their distinct emission patterns;
the SXC is probably emitted axi-symmetrically around  the star's symmetry axis,
whereas the HXC breaks that  symmetry.
Possible geometries and some inference on 
the HXC emission process are already described previously
(e.g., \cite{Makishima16}; 19; 21a),
which we do not repeat here.

\vspace*{-1mm}
\subsection{Future prospects}
\label{subsec:future_prospects}

Utilizing the earliest pulse detections from \sgr,
the present results have provided literally  the starting point 
of the  study of non-spherical dynamics of this prototypical object.
We may point out several future prospects.

One aspect of  interest with \sgr\  is whether
any noticeable change took place in $\epsilon$  or $\alpha$,
across the 2004 GF which was
presumably caused by a sudden release of internal magnetic energies
(of $B_{\rm d}$ and/or $B_{\rm t}$).
By analyzing rich  X-ray data taken after 2004 December
(e.g., with XMM-Newton, \Su\, and \NuS),
and comparing with the present results,
some information may be obtained.

The free-precession interpretation of the PPM effects predict
$T \propto P$ as in equation~(\ref{eq:slip_period}),
assuming $\epsilon$ and $\alpha$ both constant.
To verify this relation of high importance, \sgr\ is indeed ideal,
because of its rapid spin down and the shortest $T$.
This attempt will be carried out in future,
together with the above study of the source behavior across the GF.

Putting aside \sgr, another future task is to examine the suggested
time evolution of the $B_{\rm t}/B_{\rm d}$ ratio
(subsubsection~\ref{subsubsec:Bt_of_magnetars}).
For this purpose, we need to detect the PPM
and measure $T$
from  magnetars with  intermediate ages,
in between \sgr\ and the other three in table~\ref{tbl:PPM_param}.
Promising candidates include
SGR~0501+4516, 1RXS~J170849.0$-$400910,
and 	1E~1841$-$045,
which have  $\tau_{\rm c}=15$ kyr, 8.9 kyr,
and 4.6 kyr, respectively.
In fact, some preliminary information was derived affirmatively
from some of them  \citep{Makishima23}.

\section{Conclusions}
\label{sec:conclusions}
We  analyzed the 0.7--12 keV \ASCA\ GIS data of \sgr,
taken on three (effectively two) occasions  in 1993 October,
and  another in 1995 October.
The pulses, previously detected by KEA98 using the same data,
were reconfirmed,
at consistent  periods of 7.4685 s in 1993  and 7.4738 s in 1995.

In all  data sets, the 3--12 keV pulses were 
found to suffer the PPM effect,
with a modulation period of $T\approx 16.4$ ks
and an amplitude of $A \approx 1$ s.
Thus, \sgr\ becomes a fourth magnetar exhibiting this phenomenon,
with the shortest $T$,
after 4U~$0142-61$, \oneE, and SGR 1900+14.
The PPM effect must be ubiquitous among magnetars.

The PPM  in \sgr\ is present
in energies above $\sim 2.5$ keV,
but absent in lower energies.
As inferred from the previous three magnetars,
the PPM is thus associated exclusively
with their HXC, and absent in their SXC.

Using the merged 1993 data,
the PPM was confirmed to persist coherently
for a 11-d interval,
with a refined period of $16.453 \pm 0.024$ ks.
Therefore, the PPM is a highly stable phenomenon,
and its interpretation in terms of the free precession
(and the associated beat effect) is much reinforced.

Including these four examples,
all magnetars may be  axially deformed to $\epsilon \sim 10^{-4}$,
which in turn is ascribed to toroidal magnetic fields
reaching $\Bt \sim 10^{16}$ G.
All magnetars are hence suggested to harbor
toroidal magnetic fields of this strength.
In addition, $\Bt$ of magnetars may last longer than their $\Bd$.

\section*{Acknowledgements}
The present work was  supported in part by the JSPS
grant-in-aid (KAKENHI), number 21K03624.
KM thanks Dr. Hiroki Yoneda for his helpful advices.

\section*{Appendix A: The $Z_m^2$ statistics}
We explain how to estimate errors in the period determination in a $Z_m^2$ PG,
and how to distinguish among multiple period candidates seen there.
\subsection*{A1: $Z_m^2$ and the likelihood function}
Let $Z_m^2 (P)$ represents a PG calculated from an un-binned time series,
like the present case, using the $Z_m^2$ statistics.
If a particular  period $P_1$  gives a high value of $Z_m^2 (P_1) $,
we regard $P_1$ as having a high ``likelihood"
to represent the true period $P^*$.
In fact, as shown by \citet{YonedaPhD},
the so-called likelihood function ${\cal L}(P)$ (see, e.g., \cite{Likelihood} for its definition),
calculated from the same time series, satisfies a relation
\begin{equation}
Z_m^2 (P) \approx 2 \log {\cal L}(P)~,
\label{eq:AppA1}
\end{equation}
as long as the number of photons is sufficiently large,
the data are dominated by  Poisson noise,
$P$ lies in a vicinity of $P^*$,
and ${\cal L}$ around $P^*$ distributes as a Gaussian.
Note that ${\cal L}(P)$ is not a probability density,
because we are using a limited amount of data
to estimate the underlying probability density function that is unknown.

\vspace*{-5mm}

\subsection*{A2: Errors in  the period determination}
When a PG becomes highest at a period $P_0$,
it obviously provides the most likely (maximum likelihood) candidate for $P^*$.
However, we need to estimate the error associated with $P_0$\.
Hence, using equation~(\ref{eq:AppA1}),
we define an associated variable
\begin{equation}
X(P) \equiv Z_m^2(P_0) - Z_m^2(P) \approx 2 \log [{\cal L}(P_0)/{\cal L}(P)] \geq 0,
\label{eq:AppA2}
\end{equation}
which is just the {\em decrement} in $Z_m^2(P)$.
Then, the process of maximizing $Z_m^2(P)$ turns into
a process of minimizing $X(P)$.

Suppose that  the $X(P)$-minimum point $P_0$ is close to $P^*$,
and $X(P)$ is Gaussian distributed around $P_0$.
Then, as explained in \citet{YonedaPhD} and quoted by \citet{Makishima21a},
$X(P)$ obeys a chi-square distribution of $n$ degrees of freedom,
where $n$ is not the harmonic number $m$,
but the degree of freedom involved in the $X(P)$-minimization process.

A simple periodogram has $n=1$,
because  $P$ is the only variable.
The 68\%  and 90\% confidence 
error regions are specified by the condition 
that $X(P)$ reaches 1.0  and 2.71, respectively.
When the demodulation process is incorporated,
we should instead use  $n=4$ (regardless of $m$),
because the  four parameters, $P, T, A$, and $\psi$, 
are optimized to find minimum point with $X(P)=0$.
Then, the 68\% and 90\% confidence regions 
are determined by $X(P)=4.72$ and $X(P)=7.78$, respectively.
In table~\ref{tbl:results} and table~\ref{tbl:fringe_param}, 
the errors associated with $P$ and the other parameters
are determined in this way,
for both the raw and demodulated PGs,
employing the 68\% convention.

\begin{table*}
\caption{Parameters of the peaks  in the 2.55--12 keV DeMD from G93T.$^{*}$}
\label{tbl:T-fringe_param}
\begin{center}
\begin{tabular}{lcccccclll}
\hline 
DeMD  & $T$ &  $P^{ \dagger}$  &&  \multicolumn{2}{c}{$\zz$ }  &&  \multicolumn{3}{c}{Relative probability} \\
\cline{5-6} \cline{8-10} 
peak & (ks)&       (s) &&        G93T  & G95$^\ddagger$  && G93T$^\S$ & G95$^\|$& total$^\#$               \\
\hline \hline 
A & 17.520 (42) & 7.468\,140   && 62.92 & 37.22 &&0.63  & $1.8\times 10^{-4}$  & $1.2\times 10^{-4}$\\
B & 17.187 (58) & 7.468\,414   && 57.15 & 40.39 && 0.035 & $8.9\times 10^{-4}$ & $3.2\times 10^{-5}$ \\       
C & 16.782 (39)& 7.468\,482    && 56.64 & 49.23 &&0.027  & 0.074 & $2.0 \times 10^{-3}$\\
C' & 16.626 (16)& 7.468\,470    && 58.91 &  52.63 && 0.084 & 0.41 & $0.034$\\
D & 16.435 (24) &7.468\,484    && 63.86  & 54.43  &&(1)   & (1)  & (1) \\
E & 16.113 (35) &7.468\,416    && 58.41& 51.85  &&0.066 & 0.27 & 0.018\\
F & 15.807 (30) & 7.468\,417   && 59.93 & 43.66  &&0.14  & $4.6 \times 10^{-3}$ &$6.5 \times 10^{-4}$\\
G & 15.510 (27) &7.468\,416   && 57.16 & 35.56   &&0.035  &$ 7.9 \times 10^{-5}$& $ 2.8 \times 10^{-6}$\\
\hline 
\end{tabular}
\begin{footnotesize}
\begin{itemize}
\setlength{\itemsep}{0mm}
\item[$^{*}$] Referring to the red DeMD in figure~\ref{fig:G93T_DeMD}.
\item[$^\dagger$] Defined at the start of the G93pA data stream.
\item[$^\ddagger$] The $\zz$ value of the G95 DeMD in 3--12 keV.
\item[$^\S$] Corresponding to the factor $\exp(- X/2)$ in equation~(\ref{eq:AppA5}).
\item[$^\|$] Corresponding to the relative prior probability $Q(T)/Q(T_0)$ in equation~(\ref{eq:AppA5}).
\item[$^\#$] Representing the relative posterior probability $Q(T_k|X)/Q(T_0|X)$ of equation~(\ref{eq:AppA5}).
\end{itemize}
\end{footnotesize}
\end{center}
\end{table*}

\vspace*{-3mm}
\subsection*{A3: Identification of the most likely period fringe}
When a PG contains multiple peaks like figure~\ref{fig:G9395_PGs}c,
we must examine whether the most likely peak 
at $P_0$ can immediately be identified with $P^*$, 
or some other peaks must also be considered  for their candidacy.
Since $X(P)$ is no longer Gaussian-distributed,
we must abandon the method used  in subsection A2,
to regard  $X(P)$  as a variable obeying a chi-square distribution.
Instead, we may invoke the Bayes's theorem, to derive
\begin{equation}
Q(P|X) = \frac{\alpha\, {\cal L}(X|P) \times Q(P)}{Q(X)}.
\label{eq:AppA3}
\end{equation}
Here, $\alpha \ge 0$ is a  constant,
$Q(X)$ is the probability to observe the data $X \equiv X(P)$
of equation~(\ref{eq:AppA2}),
${\cal L}(X|P)$ is the likelihood of $X$  against an assumed $P$,
$Q(P)$ is the prior probability for $P$ to represent the true period $P^*$,
and $Q(P|X)$ is the posterior probability for the same statement
after the data $X$ are given.

By taking the ratio of equation~(\ref{eq:AppA3}) between  $P_0$ and another period $P$,
and substituting equation~(\ref{eq:AppA2}) 
into ${\cal L}(P)/{\cal L}(P_0) ={\cal L}(X|P)/{\cal L}(X|P_0) $,
$\alpha$ and $Q(X)$ both cancel out to give 
\begin{equation}
\frac{Q(P|X)}{Q(P_0|X)} = \frac{{\cal L}(X|P) Q(P)}{{\cal L}(X|P_0) Q(P_0)}
= \exp\left\{-\frac{1}{2}X(P)\right\} \frac{Q(P)}{Q(P_0)}.
\label{eq:AppA4}
\end{equation}
This gives a posterior probability, relative to $P_0$,
for $P$ to be the true period.

We applied equation~(\ref{eq:AppA4}) to the period fringes 
seen in the PG from G93T (figure~\ref{fig:G9395_PGs}c).
The derived posterior probability of the $n$-th fringe, 
denoted as $Q_n \equiv Q(P_n|X)$,
is given in table~\ref{tbl:fringe_param}.
Here, we set  $Q(P)/Q(P_0) \equiv 1$,
because we do not have any prior information as to the pulse period in 1993;
the results from KE98, G93p or G3f should not be used
because they are not independent  from the G93T data,
and the back extrapolation from the 1995 result is not constraining enough.
See subsection~\ref{subsec:ana_PG} for the subsequent discussion.

\subsection*{A4: Evaluation of the $T$-fringes}
The $T$-fringes seen in figure~\ref{fig:G93T_DeMD} have
parameters as summarized in table~\ref{tbl:T-fringe_param}.
Their relative significance can be examined in the same way as in subsection A3,
by  replacing $P$ in equation~(\ref{eq:AppA4}) with $T$ and  rewriting it as
\begin{equation}
\frac{Q(T_k|X)}{Q(T_0|X)} = \frac{{\cal L}(X|T_k) Q(T_k)}{{\cal L}(X|T_0) Q(T_0)}
= \exp\left[-\frac{1}{2}X(T)\right] \frac{Q(T)}{Q(T_0)}.
\label{eq:AppA5}
\end{equation}
Here, $T_0=16.435$ ks specifies the peak D, and $T_k$ any other $T$-fringe peak.
Assuming the true value of $T$ to be the same between 1993 and 1995,
we can utilize the G95 DeMD (black curve in figure~\ref{fig:G93T_DeMD})
as the prior information, 
and write  $Q(T)/Q(T_0) = \exp( -X_{\rm 95}/2)$,
where $X_{\rm 95}$ is the $\zz$ decrement in the G95 DeMD from its peak value.
The results of this calculation are summarized in the last three columns
in table~\ref{tbl:T-fringe_param},
and are utilized in subsubsection~\ref{subsubsec:demod_G93T}.

\section*{Appendix B: Significance of the DeMD peak}
Using the 3--12 keV result from G95,
we evaluate the  significance of the  DeMD peak at $\Tpk$.
Instead of performing Monte-Carlo simulations,
the actual G95 data themselves are used, after \citet{Makishima21b}.
That is,  we repeat the demodulaton computation
in the same manner as in figure~\ref{fig:G9395_DeMD_indiv}c,
but scanning $T$  from 0.05 ks to 6 ks,
which is sufficiently longer than $P$ but shorter than $\Tpk$.
Because $\Tpk$ is close to three times the orbital period of \ASCA,  
$P_{\rm orb}\approx 5.5$ ks,
the selected period range
purposely covers $P_{\rm orb} $ and $2P_{\rm orb}$.
To ensure Fourier independence between adjacent trials in $T$,
its step is chosen as $\delta T\gtrsim T^2/S$,
where $S\approx 112.4$ ks is the total data span.
We treat $A$ and $\psi$  in the same manner as in the actual PPM analysis,
and $P$ is varied by $\pm 60~\mu$s around the best  period, 7.47384 s.
The values of $Z_m^2$ thus obtained will represent 
its statistical plus systematic fluctuations 
under the actual observing condition.

The above {\it  control} study  yielded about 2100 {\it independent} trials in $T$,
but in no cases $Z_4^2$ exceeded the target value, 
$Z_4^2=54.4$ (table~\ref{tbl:results});
the highest was $Z_4^2 =51.0$.
Thus, the probability to obtain, by chance, a DeMD peak with $Z_4^2\geq 54.4$ 
in a single trial in $T$ becomes  $\Pch <1/2100=0.048\%$.
The chance  probability of the $T=16.4$ ks peak is obtained 
by multiplying this $\Pch$ with the actual number 
of {\it  independent} trials conducted in producing 
figure~\ref{fig:G9395_DeMD_indiv}c over $T=7-100$ ks.
This is estimated as $S/7-S/100 = 14.9$ 
in terms of the Fourier wave number.
These finally yield $\Pch < 0.048\% \times 14.9 = 0.72\%$.

For further confirmation,
we repeated the demodulation calculation at 
$T\approx P_{\rm orb} =16.81 \pm 0.05$ ks 
(subsubsection~\ref{subsubsec:results_individual})
and $T \approx 2P_{\rm orb} $ with a finer step with $\Delta T=0.01$ ks,
but $\zz$ remained $\leq 44.5$.
Since possible artifacts related to the ASCA's orbital period is
thus signifiant at neither $T \approx P_{\rm orb} $ 
nor $T \approx 2P_{\rm orb}$,
we reconfirm that
the DeMD peak at $T=\Tpk \approx 3P_{\rm orb}$ is
very unlikely to be instrumental.

From these evaluations, we can claim, 
at  99\% or higher confidence,
that the DeMD peak at $\Tpk$ in the 3--12 keV GIS95 data is real.


\begin{thebibliography}{99}
\bibitem[Appourchaux et al.(1998)]{Likelihood} 
Appourchaux, T., L. Gizon, L., \&  Rabello-Soares, M.-C. 1998,
\aap\ Suppl., 132, 107
\bibitem[Burke et al.(1991)]{SIS} 
Burke, B. E., et al. 1991, IEEE Trans., ED38, 1069
\bibitem[Butikov(2006)]{Butikov06} 
Butikov, E. 2006,  Europ. J. Phys.,  27, 1071
\bibitem[Enoto et al.(2011)]{Enoto11} 
Enoto, T., et al. 2011,
PASJ, 63,  387
\bibitem[Enoto et al.(2010)]{Enoto10} 
Enoto, T., Nakazawa, K.,  Makishima, K., Rea, N.,  Hurley, K.,  \& Shibata, S.  2010, 
\apjl, 722, L162
\bibitem[Ioka \& Sasaki(2004)]{Ioka&Sasaki04} 
Ioka, K.,  \& Sasaki, M. 2004,  \apj, 600, 296
\bibitem[Kouveliotou et al.(1994)]{Kouveliotou94} 
Kouveliotou, C., et al. 
1994, \nat, 368, 125
\bibitem[Kouveliotou et al.(1998)]{Kouveliotou98} 
Kouveliotou, C., et al. 
1998, \nat, 393, 235 (KEA98)
\bibitem[Landau and Lifshitz(1996)]{Landau&Lifshitz} 
Landau, L. D., \&  Lifshitz, E. M. 1996, Mechanics (3rd.ed). Pergamon
\bibitem[Makishima (2023)]{Makishima23} 
Makishima, K. 2023, Proc. IAU, 363, 267
\bibitem[Makishima et al.(1996)]{GIS2} 
Makishima, K., et al. 1996, \pasj, 48,171
\bibitem[Makishima et al.(2014)]{Makishima14} 
Makishima, K., Enoto, T., Hiraga, J. S., Nakano, T., 
Nakazawa, K., Sakurai, S., Sasano, M., \& Murakami, H. 2014, 
Phys. Rev. Lett., 112, 171102
\bibitem[Makishima et al.(2016)]{Makishima16}  
Makishima, K., Enoto, T.,  Murakami, H., Furuta, Y., 
Nakano, T., Sasano, M., \& Nakazawa, K. 2016, 
PASJ,  68, S12
\bibitem[Makishima et al.(2021a)]{Makishima21a} 
Makishima, K.,  Enoto, T., Yoneda, H., \& Odaka, H.  2021a, 
\mnras,  502, 2266
\bibitem[Makishima et al.(2019)]{Makishima19}  
Makishima, K., Muakami, H., Enoto, T., \& Nakazawa, K.  2019,
PASJ, 71, 15
\bibitem[Makishima et al.(2021b)]{Makishima21b} 
Makishima, K., Tamba, T., Aizawa, Y.,  Odaka, H., 
Yoneda, H., Enoto T.,  \& Suzuki, M.  2021b, 
\apj,  923, 63
\bibitem[Makishima et al.(2023)]{LS5039GIS} 
Makishima, K., Uchida, N., Yoneda, H., Enoto T.,  \& Takahashi, T.  2023,
\apj,  959,  id.79
\bibitem[Mereghetti(2008)]{Mereghetti08} 
Mereghetti, S. 2008, A\&AR, 15, 225
\bibitem[Murakami et al.(1994)]{Murakami94} 
Murakami, T.,  Tanaka, Y., Kulkarni, S., Ogasaka, Y., 
Sonobe, T., Ogawara, Y., Aoki T.,  \& Yosida A.
1994,  \nat, 368, 127
\bibitem[Nakano et al.(2015)]{Nakano15} 
Nakano, T., Murakami, H., Makishima, K., Hiraga, J. S., 
Uchiyama, H., Kaneda, H., \& Enoto, T. 2015, \pasj, 67, id,9
\bibitem[Ohashi et al.(1996)]{GIS1} 
Ohashi, T., et al. 1996, \pasj, 48, 157
\bibitem[Palmer et al.(2005)]{GiantFlare}
Palmer, D. M., et al. 2005,
\nat,  434, 1107.
\bibitem[Rea et al.(2014)]{Rea14}
Rea, N., Vigan\'o,  D.,  Israel, G. L.,  Pons, J. A., \& Torres, D. F. 2014, 
\apjl,  781, L17.
\bibitem[Serlemitsos et al.(1995)]{XRT}
Serlemitsos, P., et al. 1995, \pasj, 47, 105
\bibitem[Tanaka et al.(1994)]{Tanaka94} 
Tanaka, Y., Inoue, H.,  \&  Holt, S. S. 1994, PASJ,  46, L37
\bibitem[Younes et al.(2015)]{Younes15}
Younes, G., Kouveliotou, C.,  \& Kaspi, V. M.
2015, \apj, 809, id. 165
\bibitem[Yoneda(2020)]{YonedaPhD}
Yoneda, H. 2020, PhD Thesis, The University of Tokyo,
Appendix A
\end{thebibliography}
\end{document}